\documentclass[lettersize,journal]{IEEEtran}
\usepackage{amsmath,amsfonts}
\usepackage{algorithmic}
\usepackage{algorithm}
\usepackage{array}
\usepackage{newtxtext}
\usepackage[colorlinks,citecolor=blue,linkcolor=blue,urlcolor=blue]{hyperref}
\usepackage[caption=false,font=footnotesize,labelfont=rm,textfont=rm]{subfig}
\usepackage{textcomp}
\usepackage{stfloats}
\usepackage{url}
\usepackage{verbatim}
\usepackage{graphicx}
\usepackage{cite}
\usepackage{bm}
\usepackage{cuted}
\usepackage{amssymb}
\usepackage{color}

\begin{document}

\title{FAS-RSMA: Can Fluid Antennas Elevate RSMA Performance?}

\author{Jinyuan~Liu,
        Yong~Liang~Guan,~\IEEEmembership{Senior~Member,~IEEE,}
        Tuo~Wu,~\IEEEmembership{Member,~IEEE,}\\
        Kai-Kit~Wong,~\IEEEmembership{Fellow,~IEEE,}
        and~Bruno~Clerckx,~\IEEEmembership{Fellow,~IEEE}
\thanks{Jinyuan Liu and Yong Liang Guan are with the School of Electrical and Electronic Engineering, Nanyang Technological University, Singapore 639798 (e-mail:
jinyuan001@e.ntu.edu.sg; eylguan@ntu.edu.sg).}
\thanks{Tuo Wu is with the Department of Electrical Engineering City University of Hong Kong, Hong Kong, China. (E-mail: tuo.wu@qmul.ac.uk).}
\thanks{K. K. Wong is with the Department of Electronic
and Electrical Engineering, University College London, London, UK. (e-mail: {kai-kit.wong}@ucl.ac.uk). K. K. Wong is also affiliated with Yonsei Frontier Lab, Yonsei University, Seoul, Korea.
}
\thanks{Bruno Clerckx is with the Department of Electrical and Electronic Engineering, Imperial College London, London SW7 2AZ, U.K. (e-mail: b.clerckx@imperial.ac.uk).}
}



\maketitle
\setlength{\abovedisplayskip}{4.0pt}
\setlength{\belowdisplayskip}{4.0pt}

\begin{abstract}
As sixth-generation (6G) wireless networks demand unprecedented connectivity and interference management capabilities, rate-splitting multiple access (RSMA) emerges as a promising solution through common and private stream partitioning and remains effective across a range of channel state information at the transmitter (CSIT) qualities and traffic heterogeneity. In practical multiuser deployments, two considerations arise: the common stream decoding constraint imposed by the weakest user, and residual inter-user interference can remain non-negligible—particularly in single-input single-output (SISO) broadcast settings and under an imperfect CSIT scenario. Motivated by prior advances of RSMA research, we investigate a complementary mechanism—fluid antenna systems (FAS) with dynamic port reconfiguration—that supplies adaptive spatial selectivity without altering the RSMA signaling structure. Can FAS help alleviate these considerations and enhance RSMA performance? This paper demonstrates that dynamic port reconfiguration in FAS provides adaptive spatial selectivity that can strengthen the weakest user’s effective channel, improve signal-to-interference-plus-noise (SINR) ratios through enhanced channel gains and reduced relative noise impact. We develop a tractable, correlation-aware analytical framework that captures realistic spatial dependencies through advanced block-correlation modeling, considering both constant block correlation (CBC) and variable block correlation (VBC) variants. Our analysis yields closed-form expressions for outage probability (OP) and average capacity (AC) that quantify the impact of FAS on RSMA performance. Extensive simulations validate our theoretical findings: VBC-based results exhibit consistently tighter agreement with Monte Carlo simulations than CBC across all port configurations. Moreover, FAS-RSMA achieves enhancing performance gains over traditional antenna systems (TAS) and non-orthogonal multiple access (NOMA), demonstrating lower OP and substantially higher AC through the synergy of RSMA's flexible interference management and FAS's adaptive spatial diversity.  
\end{abstract}


\section{Introduction}
Sixth-generation (6G) wireless systems are expected to deliver step-change improvements in throughput, reliability, coverage, and connection density, thereby enabling immersive applications, ultra-reliable low-latency communications (URLLC), integrated sensing and communications (ISAC), and AI-native services \cite{ref1}. To achieve these ambitious targets, 6G networks must support massive connectivity with heterogeneous quality-of-service (QoS) requirements while efficiently managing interference in dense deployment scenarios, where traditional orthogonal multiple access schemes face severe limitations in spectral efficiency and flexibility \cite{ref2}. To address this fundamental challenge of multiuser interference, rate-splitting multiple access (RSMA) has emerged as a promising solution \cite{ref20,ref21,ref210,ref2105,ref211,ref212,ref22,ref221,ref23,ref232,ref233,ref234,ref24}. RSMA partitions each user's information into a \emph{common} stream—decoded by all users—and a \emph{private} stream—decoded only by the intended user, thereby providing a continuous tradeoff between fully decoding interference and treating interference as noise. Extensive performance analysis research of RSMA across diverse settings, including UAV-assisted networks \cite{ref241}, physical-layer security \cite{ref243}, and semi-grant-free systems \cite{ref244}, consistently demonstrates RSMA's superior reliability and flexibility compared with space-division multiple access (SDMA) and non-orthogonal multiple access (NOMA
While RSMA is a flexible and robust multiple-access scheme, two practical constraints require consideration due to their potential impact on the achievable performance of RSMA. The first one is the \emph{common stream decoding}: since all users must successfully decode the common stream, the user with the poorest channel condition becomes the performance bottleneck, constraining the achievable common stream rate and overall system throughput. The second is the \emph{residual inter-user interference}: although RSMA's rate-splitting strategy provides flexible interference management, residual interference from private streams of other users still degrades the signal-to-interference-plus-noise ratio (SINR) performance, which is particularly pronounced under single-input single-output (SISO) multi-user scenarios and imperfect channel state information (CSI) conditions. To address these issues, representative extensions have been proposed including hierarchical rate-splitting (HRS) \cite{ref245}, which layers multiple common streams to better match channel and traffic heterogeneity, and cooperative rate-splitting (CRS) \cite{ref246}, which enables a user to forward the decoded common stream to assist others.

Taken together, HRS and CRS have advanced RSMA’s performance and robustness. Motivated by this progress, we ask whether a physical, hardware-centric mechanism—fluid antenna systems (FAS) \cite{ref4,ref5,ref18,ref26,ref27,ref28,ref6,ref7,ref8,ref84,ref85,ref9,ref10,ref11,ref110,ref12,ref13,ref14,ref15,ref16,ref17,ref19} with dynamic port reconfiguration—can offer a complementary approach to address the aforementioned considerations. Unlike conventional fixed-position arrays, an FAS allows the radiating element (or an equivalent port) to be repositioned within a prescribed aperture—either physically or virtually—so that the terminal can opportunistically ``sample" the spatial channel at multiple locations \cite{ref18}. Early prototypes investigated fluidic implementations utilizing conductive liquids within precision-controlled structures \cite{ref26}. However, technologies such as reconfigurable metasurfaces \cite{ref27} and spatially distributed antenna pixels \cite{ref28} are more transformative by eliminating any response time for the FAS reconfigurability. This reconfigurability represents a paradigm shift from fixed-position antenna (FPA) arrays, in which element locations remain permanently static. Dynamic control of spatial points as radiating elements in FASs introduces spatial degrees of freedom (DoFs), resulting in a virtual aperture that transcends the constraints of conventional antenna arrays. This dynamic reconfiguration introduces an additional physical-layer degree of freedom that is, to a large extent, decoupled from the RF chain budget. By switching among a dense set of candidate ports along a flexible medium, FAS can help elevate RSMA's performance: \emph{first}, by enabling each user to select the port with maximum channel gain, FAS strengthens the channel quality of the weakest user, thereby alleviating the weakest-user that bounds the common-stream rate; \emph{second}, through spatial selectivity, FAS allows users to choose positions with maximum channel gain where both desired and interfering signals are enhanced, but the desired signal benefits more from dedicated power allocation while the relative impact of fixed noise power diminishes, enhancing inter-user interference management through improved SINR ratios. Recent studies have begun to formalize FAS-RSMA architectures: \cite{ref250} studied an FAS-aided RSMA-integrated sensing and communication (ISAC) system; \cite{ref25} developed a downlink FAS-RSMA framework; \cite{ref29} investigated STAR-RIS assisted RSMA with FAS users; and \cite{ref30} analyzed UAV-relay-assisted RSMA networks with FAS users.

While significant progress has been made in FAS-RSMA performance analysis, the analytical characterization of this combined system presents multi-layered challenges that motivate further research. The fundamental complexity stems from RSMA's sophisticated two-stage decoding process, where users first decode a common stream while treating interference as noise, followed by successive interference cancellation (SIC) and private stream decoding. This intricate decoding structure inherently complicates the derivation of tractable performance expressions, as the interdependencies between common and private stream decoding create complex analytical relationships. The integration of FAS introduces additional layers of complexity through two key factors: the dynamic nature of port selection, which involves optimization over multiple spatial positions, and the spatial correlation among channel coefficients across different ports. The dynamic port reconfiguration capability, while providing performance benefits, makes the system behavior highly dependent on the joint statistics of spatially correlated channels, further complicating closed-form analysis. Existing studies have made valuable contributions by deriving OP expressions using the multivariate $t$-distribution \cite{ref25,ref29,ref30} and copula-based formulations \cite{ref29}, though these approaches often require computationally intensive numerical evaluations or exhibit limited agreement with Monte Carlo simulations under diverse operating conditions. The channel covariance encountered in FAS typically exhibits a Toeplitz (or Toeplitz-like) structure under wide-sense stationary scattering assumptions \cite{ref18}, adding another dimension of analytical complexity. Prior work has made important contributions by adopting \emph{block-correlation} models \cite{ref18,ref19}, with conventional constant block-correlation (CBC) modeling providing valuable insights for large port configurations. However, when only a moderate number of ports is available, individual ports exert stronger influence on second-order statistics \cite{ref2,ref6,ref9,ref12}, highlighting the need for more flexible modeling approaches. The original work in \cite{ref18} recognized this limitation and proposed the concept of variable block-correlation modeling (VBC), but the induced high-complexity optimization precluded \cite{ref18} from providing an explicit algorithm for optimal parameter selection. To bridge this gap and enable practical use of VBC, recent work has developed an algorithmic framework for systematically selecting optimal block parameters \cite{ref17}; however, its application to FAS–RSMA performance analysis remains unexplored. To enable rigorous quantitative analysis and provide deeper insights into how FAS enhances RSMA's performance characteristics, there is a compelling need for more tractable analytical frameworks that can facilitate practical system design and optimization.

To address these challenges and provide a rigorous analytical framework for quantifying how FAS elevates RSMA's performance, this paper introduces advanced block-correlation modeling to the SISO multi-user FAS-RSMA downlink communication system. By applying sophisticated mathematical tools and adopting both CBC and VBC models to capture spatial correlation accurately, we derive closed-form expressions for key performance metrics, including OP and average capacity (AC). Our analysis identifies mechanisms by which FAS-enabled port-selection diversity can alleviate the weakest-user coupling in common-stream decoding and elevate inter-user interference management, providing crucial insights for practical system design. Specifically, this paper makes the following key contributions:

\begin{itemize}
    \item \textbf{Enhanced FAS-RSMA Analysis Framework:} We develop a tractable performance–analysis framework for a downlink multiuser FAS-RSMA system that explicitly accounts for spatial correlation via block–correlation modeling. Beyond the conventional CBC model assumption, the proposed framework adopts a VBC model in which the intra–block correlation coefficients are allowed to differ across blocks, thereby capturing spatial heterogeneity and yielding substantially higher accuracy than CBC.

    \item \textbf{Rigorous Mathematical Derivation:} We derive novel closed-form expressions for the users' OP and for the AC of both common and private streams within a mathematically consistent analytical framework. In contrast to copula-based formulations that typically lead to multivariate $t$-distributions and heavy numerical evaluation, the proposed closed-form expressions enable fast and accurate performance assessment across system parameters without resorting to computationally intensive numerical integration.

    \item \textbf{Extensive Performance Validation:} Extensive simulations across diverse configurations corroborate the theoretical analysis. VBC-based evaluation exhibits markedly tighter agreement with Monte Carlo benchmarks than CBC, indicating a more accurate characterization of the spatial correlation structure among the channel coefficients. Meanwhile, RSMA consistently achieves lower OP than NOMA under a variety of FAS settings of near-equal power splits of the users: RSMA's common-stream design converts shared power into interference suppression and reliability gains, whereas NOMA's successive decoding under residual interference yields poorer OP. Moreover, compared with traditional antenna RSMA systems (TAS-RSMA), FAS-RSMA achieves additional gains in both OP and AC, attributable to the port–selection diversity and dynamic port reconfiguration afforded by FAS.
    
    \item \textbf{Practical Design Insights:} By demonstrating substantial improvements in both OP and AC relative to TAS-RSMA, our analysis makes a compelling case for jointly leveraging FAS and RSMA in future 6G systems. In addition, we provide practical design guidelines for FAS-RSMA that cover power-allocation strategies and multiple-access configurations, thereby facilitating deployment in real-world systems.
\end{itemize}

\textit{Notation:} Matrices are denoted by boldface uppercase letters, column vectors are denoted by boldface lowercase letters, and scalars are denoted by standard letters. $x \sim \mathcal{C} \mathcal{N}\left(\alpha, \beta\right)$ denotes a complex Gaussian random variable with mean $\alpha$ and variance $\beta$.  $\mathbb{E}\left \{\cdot \right \}$ denotes the expectation of an random variable. $f_X \left(x \right)$ and $F_X\left(x \right)$ denote the probability density function (PDF) and cumulative distribution function (CDF), respectively. $\left \| \cdot   \right \| _2$ represents Euclidean norm.

The remainder of this paper is organized as follows. Section~II details the FAS-RSMA system model and the adopted block–correlation models. Section~III develops the performance analysis and derives closed-form expressions for OP and AC. Section~IV reports numerical experiments and validates the analysis against Monte Carlo simulations. Finally, Section~V concludes the paper and outlines directions for future research.
\vspace{-10pt}
\section{System Model for FAS-RSMA}
\begin{figure}[ht]
\setlength{\belowcaptionskip}{-0.9cm} %
\vskip 0.1in
\begin{center}
\centerline{\includegraphics[width=0.52\textwidth]{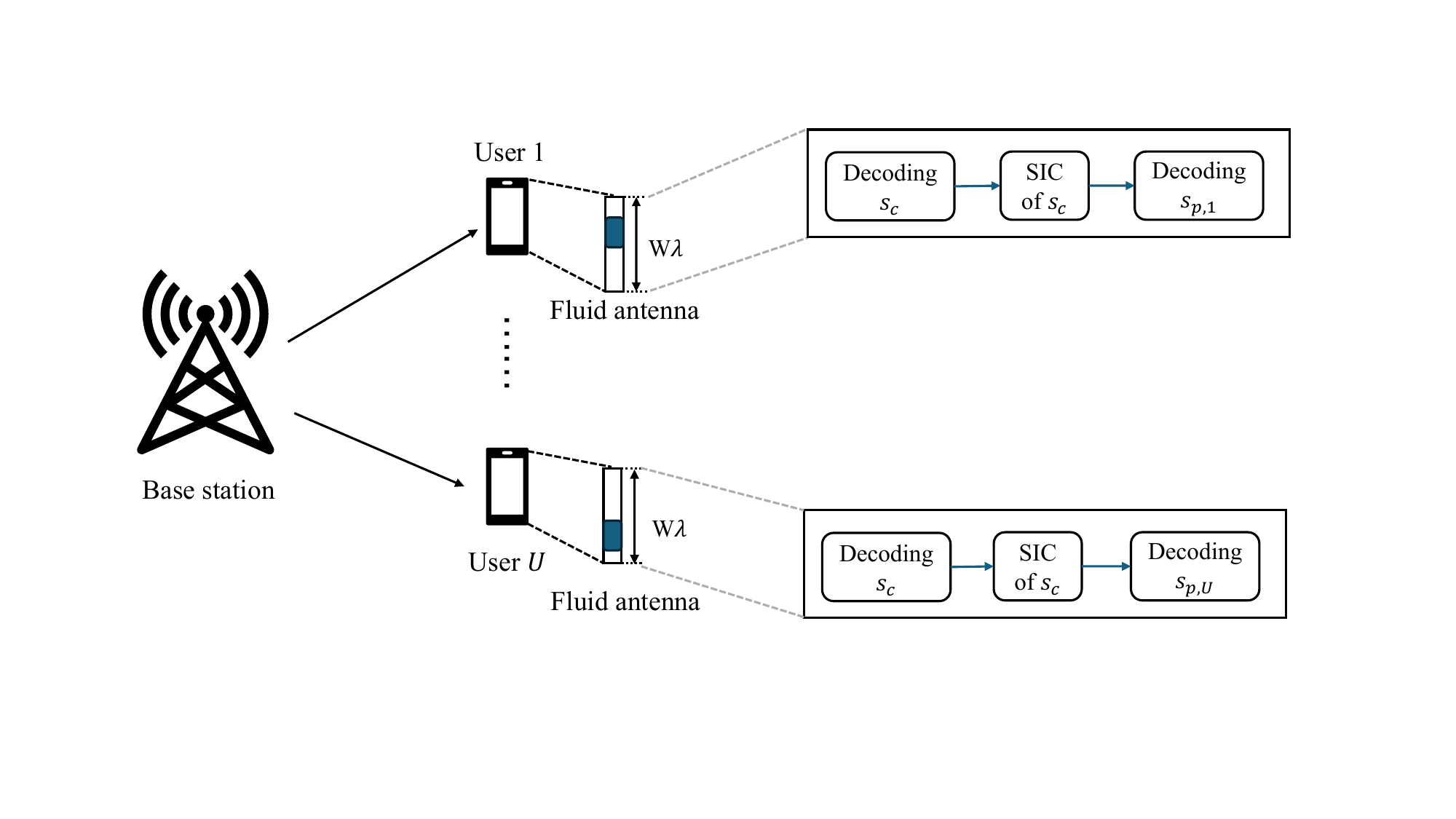}}
\caption{Multi-user downlink FAS-RSMA communication system.}
\vspace{-20pt}
\label{fig:System model}
\end{center}
\vskip -0.1in
\end{figure}

We consider a downlink communication system where a base station (BS) equipped with a traditional single-antenna system (TAS) serves $U$ users simultaneously. All users are equipped with one-dimensional linear fluid antennas that enable switching among $N$ discrete ports \cite{ref17,ref19,ref25}. These ports are distributed within a linear space of length $W \lambda$, where $W$ is a normalization parameter and $\lambda$ is the carrier wavelength. The proposed FAS-RSMA system model is illustrated in Fig. \ref{fig:System model}.
\vspace{-10pt}

\subsection{Signal Model}
At the transmitter, each user's message is split into common and private parts. All common-part messages are grouped and encoded into a common stream, denoted as $s_{c}$. All private messages are encoded independently into private streams $s_{p,u}$ for each user $u$. The overall transmitted signal at the BS can be expressed as follows:
\begin{equation}
  {s}= \sqrt{P}  \left ( \sqrt{t_c} {s}_{c}  +  \sum_{u=1}^{U} \sqrt{t_u}  {s}_{p,u} \right ) , 
\label{RSMA_transmit_signal} 
\end{equation}
where $P$ denotes the total transmission power, $t_c$ and $t_u$ represent the power allocation factors for the common stream and the $u$-th user's private stream, respectively. $s_c$ and $s_{p,u}$ are the common and private data streams, respectively. To meet the total power constraint, we have $t_c+\sum_{u=1}^{U} t_u=1$. Then, the received signal of the $u$-th user at the $n$-th port\footnote{ Note that estimating CSI at every port is not required. Measuring a subset of locations and exploiting spatial correlation, channel sparsity, and other implicit structural properties enables reconstruction of the channel across the entire fluid-antenna aperture \cite{ref11,ref110}. } is defined as
\begin{equation}
  {y}_{n,u}= \sqrt{P t_c} h_{n,u} {s}_{c}  +   \sum_{u=1}^{U} \sqrt{Pt_u} h_{n,u} {s}_{p,u} + \omega_{n,u}, 
\label{RSMA_recevied_signal} 
\end{equation}
with $h_{n,u} \sim \mathcal{C} \mathcal{N}\left(0, \eta_0\right)$ being the channel coefficient from the BS to the $n$-th port of $u$-th user. $\omega_{n,u}$ is the independent and identically distributed (iid) additive white Gaussian noise (AWGN) with variance $\sigma^{2}$. Following the FAS concept, we assume that the optimal port with the maximum channel gain is activated, which can be written as follows:
\begin{equation}
    n^{*}=\arg \max _{n \in[1, N]} g_{n,u},
    \label{select}
\end{equation}
where $g_{n,u}=\left | h_{n,u}  \right |^2 $. Hence, $g_{n^{*},u}$ denotes the maximum channel gain of the selected port. For notational simplicity, we henceforth suppress the index $n^{*}$ and write $g_{u} \triangleq g_{n^{*},u}$, with the dependence on the selected port understood from context. Thus, the equivalent channel gain for the $u$-th user is defined as
\begin{equation}
    g_{u}= \max \left\{g_{1,u}, \dots , g_{N,u}   \right \}.
    \label{eq:channel_gain}
\end{equation}

Crucially, this port selection process operates on instantaneous CSI, enabling real-time adaptation to channel variations without requiring channel prediction or extrapolation from outdated estimates. Unlike conventional fixed-antenna RSMA systems that must rely on potentially stale CSI for power allocation and scheduling decisions, FAS-RSMA can exploit the current channel realization directly, thereby mitigating the performance degradation typically associated with CSI estimation errors, feedback delays, and channel prediction inaccuracies.

Based on the RSMA principle, the decoding process for all users consists of two stages. First, each user decodes the common stream $s_c$ by treating other signals as noise. Upon successfully decoding the common symbols, successive interference cancellation (SIC) is applied to eliminate the common stream, and the $u$-th user then decodes its private stream ${s}_{p,u}$ by treating the remaining interference as noise in the second stage. The expressions for the common signal-to-interference-plus-noise ratio (SINR) $\Gamma_{c,u}$ and private SINR $\Gamma_{p,u}$ are given below:
\begin{subequations}\label{SINRs}
\begin{align}
 &\Gamma_{c,u}=\frac{\bar{\gamma } t_c  g_{u}   }{\bar{\gamma } \left ( 1-t_c \right ) g_{u}  +1 }, \\
 &\Gamma_{p,u}=\frac{\bar{\gamma } t_u  g_{u}   }{\bar{\gamma } \left ( 1-t_c-t_u  \right ) g_{u} +1 },
\end{align}
\end{subequations}
with $\bar{\gamma }=\frac{P}{\sigma^{2}}$ represents the average transmit signal-to-noise ratio (SNR). 

\vspace{-15pt}
\subsection{FAS Channel Correlation Model}
Due to the limited spacing between neighboring antenna ports, the channel responses exhibit spatial correlation, which cannot be neglected in the performance analysis. Let the channel vector be denoted as $\mathbf{h}_{u}=\left [h_{1,u}, \ldots, h_{N,u} \right ] ^{T}$, with its spatial correlation characterized by the covariance matrix $\bm{\Sigma}_{u}= \mathbb{E} \left [ \mathbf{h}_{u} \mathbf{h}_{u}^{H} \right ] $. According to the Jakes fading model, the spatial correlation coefficient between the $k$-th and $l$-th antenna ports can be modeled as
\begin{equation}
    \sigma_{k, l}=J_{0}\left(\frac{2 \pi|k-l| W}{N-1}\right),
    \label{Jakes}
\end{equation}
where $J_{0}$ denotes the zeroth-order Bessel function of the first kind. Consequently, the spatial covariance matrix $\bm{\Sigma}_{u}$ can be expressed as
\begin{equation}
\begin{aligned}
\boldsymbol{\Sigma}_{u} \in \mathbb{R}^{N \times N} & =\textbf{toeplitz}\left(\sigma_{1,1}, \sigma_{1,2}, \cdots, \sigma_{1, N}\right) \\
& =\left(\begin{array}{cccc}
\sigma_{1,1} & \sigma_{1,2} & \cdots & \sigma_{1, N} \\
\sigma_{1,2} & \sigma_{1,1} & \cdots & \sigma_{1, N-1} \\
\vdots & & \ddots & \vdots \\
\sigma_{1, N} & \sigma_{1, N-1} & \cdots & \sigma_{1,1}
\end{array}\right),
\end{aligned}
\label{eq:toeplitz_matrix}
\end{equation}
which exhibits a Toeplitz and Hermitian structure, reflecting the uniform spatial correlation determined by the relative positions of the antenna ports. Due to the Toeplitz covariance structure, analytical derivation of the channel statistics is nontrivial and often precludes closed-form expressions.

\section{Performance Analysis for FAS-RSMA}
With the system model incorporating spatially correlated wireless channels established, we proceed to investigate the performance of FAS-RSMA communication networks. In particular, our focus is on quantifying the impact of spatial correlation on the outage probability (OP) and average capacity (AC).

The performance analysis of the FAS-RSMA system proceeds in several steps. First, we formalize the core performance indicators that serve as evaluation metrics. Subsequently, the block-correlation model is employed to effectively capture and characterize the intricate spatial correlation patterns observed in FAS systems. Building upon this foundation, we derive closed-form analytical expressions for the OP and AC.
\vspace{-15pt}

\subsection{Performance Metrics}
The OP is a key performance metric in wireless communication systems that characterizes the likelihood that the instantaneous SINR falls below a predefined threshold, resulting in communication failure. In the context of the FAS-RSMA framework, each user receives a superposition of three components: a common message intended for all users, its own private message, and the interfering private messages of other users. Decoding is performed through a successive two-stage procedure. An outage event occurs if the SINR associated with either stage does not satisfy the corresponding quality-of-service (QoS) requirements \cite{ref212}. Given the SINR thresholds $\gamma^{c,u}_{\mathrm{th}}$ and $\gamma^{p,u}_{\mathrm{th}}$ for common and private streams, respectively, we establish the following proposition. \\
\textbf{Proposition 1.} \textit{The OP of the FAS-RSMA user can be defined as follows:
\begin{equation}
    P_{out}^{u}=F_{g_{u}} \left (\gamma^{u}_{\mathrm{th}}  \right ),
    \label{eq:P1}
\end{equation}
where $F_{g_{u}} \left ( \cdot  \right ) $ denotes the CDF of the maximum channel gain of the selected port, with $\gamma^{u}_{\mathrm{th}}=\max \left \{ \tilde{\gamma}^{c,u}_{\mathrm{th}}, \tilde{\gamma}^{p,u}_{\mathrm{th}} \right \}$, and the corresponding $ \tilde{\gamma}^{c,u}_{\mathrm{th}}$ and $\tilde{\gamma}^{p,u}_{\mathrm{th}}$ are defined as
\begin{subequations}\label{gamma}
\begin{align}
 &\tilde{\gamma}^{c,u}_{\mathrm{th}}=\frac{\gamma^{c,u}_{\mathrm{th}}}{\bar{\gamma } \left (t _c-\sum _u^{U}t_{u} \gamma^{c,u}_{\mathrm{th}}  \right ) } 
    , \\
 &\tilde{\gamma}^{p,u}_{\mathrm{th}}=\frac{\gamma^{p,u}_{\mathrm{th}}}{\bar{\gamma } \left (t_u-\sum _{{u}'\ne u} ^{U}t _{{u}'} \gamma_{p,{u}'}  \right ) }.
\end{align}
\end{subequations}
}

\textit{Proof.} Based on the definition of the OP, an outage event for user $u$ is declared when the decoding of either the private stream or the common stream fails to meet the required SINR threshold, i.e.,
\begin{align}\label{p1}
P_{\mathrm{out}}^{u}= & \mathrm{Pr} \bigg(\frac{\bar{\gamma } t_c  g_{u}   }{\bar{\gamma } \left ( 1-t_c \right ) g_{u}  +1 }< \gamma^{c,u}_{\mathrm{th}}, \notag\\
&\frac{\bar{\gamma } t_u  g_{u}   }{\bar{\gamma } \left ( 1-t_c-t_u  \right ) g_{u}  +1 } < \gamma^{p,u}_{\mathrm{th}}  \bigg)\notag\\
=& \mathrm{Pr} \left (g_{u} <\tilde{\gamma}^{c,u}_{\mathrm{th}}, g_{u} <\tilde{\gamma}^{p,u}_{\mathrm{th}}  \right ) \notag\\
=& \mathrm{Pr}\left ( g_{u}\le \max \left \{ \tilde{\gamma}^{c,u}_{\mathrm{th}},\tilde{\gamma}^{p,u}_{\mathrm{th}} \right \}  \right ) \notag\\
=& F_{g_{u}} \left (\gamma^{u}_{\mathrm{th}}  \right ),
\end{align} 
which proves the stated result.\\
\textbf{Remark 1:} \textit{To ensure that (\ref{gamma}) is valid and non-negative, the thresholds must satisfy the following constraints:
\begin{subequations}\label{constraints}
\begin{align}
&\tilde{\gamma}^{c,u}_{\mathrm{th}}<\frac{t_{c}}{\sum_{u=1}^{U} t_{u}}, \forall u, \\
&\tilde{\gamma}^{p,u}_{\mathrm{th}}<\frac{t_{u}}{\sum _{{u}'\ne u} ^{U}t _{{u}'}}, \forall u,
\end{align}
\end{subequations}
which means that if thresholds exceed these bounds, the algebraic SINR in \eqref{gamma} becomes nonphysical (negative or undefined), which corresponds to an infeasible decoding requirement. In this case, OP occurs with probability one, and the analytical
expressions lose validity.} 

Given the common and private SINR expressions in (\ref{SINRs}), we can write the achievable common and private rates as
\begin{subequations}\label{eq:datarate}
\begin{align}
 & R_{c, u}=\log_{2}(1+\Gamma_{c,u}), \\
 & R_{p, u}=\log_{2}(1+\Gamma_{p,u}).
\end{align}
\end{subequations}
Then the total sum rate can be defined as
\begin{equation}\label{totalrate}
    R= R_{c, u^{*}} + \sum^{u}_{u=1}  R_{p, u},
\end{equation}
with $u^{*}=\arg \min_{u} R_{c, u}$, which implies that the user associated with the smallest selected channel amplitude $g_{u^{*}}$ is chosen, which guarantees that all the users can decode the common message. Based on this, we establish the following proposition regarding the AC of the FAS-RSMA system.\\
\textbf{Proposition 2.} \textit{The AC of the FAS-RSMA system can be defined as follows:
\begin{equation}\label{eq:AC}
\bar{C}\!=\!\!\!\int_{0}^{\infty }\!\!\!  R_{c, u^{*}} f_{g_{u^{*}} }  \left ( x \right )\mathrm{d}x+ \sum^{u}_{u=1}\!\int_{0}^{\infty }\!\!\!  R_{p, u} f_{g_{u}}\left ( x \right )  \mathrm{d}x, 
\end{equation}
where $f_{g_{u^{*}} } $ and $f_{g_{u}}$ denote the PDF of the smallest selected channel amplitude among the users and the maximum channel amplitude of the $u$-th user, respectively. The above expression is obtained by computing the expected value of the total transmission rate, i.e., $\mathbb{E}\left \{ R \right \}$. }

Based on \textbf{Propositions 1} and \textbf{2}, it becomes essential to characterize the statistical distributions of the maximum channel amplitude $g_{u}$ for the $u$-th user and $g_{u^{*}}$ for the selected channel amplitude of user $u^{*}$. Nevertheless, obtaining an exact analytical expression is non-trivial due to the intricate dependencies introduced by the spatial correlation among the channel ports, which leads to highly structured and non-diagonal correlation matrices.
\vspace{-10pt}

\subsection{Block-Correlation Model}
To address the aforementioned challenge, we apply the block-correlation model initially proposed in \cite{ref18}, which provides an efficient approach by partitioning correlated channel coefficients into several independent blocks. The block-correlation model approximates the spatial covariance matrix $\bm{\Sigma}_{u}$ as a block-diagonal matrix defined as follows:
\begin{equation}
\widehat{\boldsymbol{\Sigma}}_u \in \mathbb{R}^{N \times N} =
\left[
\begin{array}{cccc}
\mathbf{A}_1 & \mathbf{0} & \cdots & \mathbf{0} \\
\mathbf{0}   & \mathbf{A}_2 & \cdots & \mathbf{0} \\
\vdots       & \vdots       & \ddots & \vdots \\
\mathbf{0}   & \mathbf{0}   & \cdots & \mathbf{A}_D
\end{array}
\right],
\label{eq:block_matrix}
\end{equation}
where each block matrix $\mathbf{A}_d \in \mathbb{R}^{L_d \times L_d}$ for $d=1, \ldots, D$ is given by
\begin{equation}
\mathbf{A}_d \in \mathbb{R}^{L_d \times L_d} =
\left[
\begin{array}{cccc}
1     & \rho_d & \cdots & \rho_d \\
\rho_d & 1     & \cdots & \rho_d \\
\vdots & \vdots & \ddots & \vdots \\
\rho_d & \rho_d & \cdots & 1
\end{array}
\right],
\label{eq:block_matrix_single}
\end{equation}
Here, $D$ denotes the number of sub-blocks, $\rho_d \in \left [ 0,1  \right ]$ represents the correlation coefficient of the $d$-th block, and $\sum_{d=1}^{D} L_d=N$, where $L_d$ defines the size of the $d$-th block. It can be observed that the underlying principle of this block-correlation approximation is to partition the original correlated channels into $D$ independent blocks, where each block shares a common correlation coefficient. To ensure that the block-correlation model provides an accurate approximation of the covariance matrix, the parameters $D$, $\left \{\rho_d  \right \}_{d=1}^{D}$, and $\left \{L_d  \right \}_{d=1}^{D}$ should be optimized by minimizing the following objective function:
\begin{equation}
\arg \min_{D, \left \{\rho_d  \right \}_{d=1}^{D}, \left \{L_d  \right \}_{d=1}^{D}} \; \mathrm{dist}\big( \boldsymbol{\Sigma}_h, \widehat{\boldsymbol{\Sigma}}_h \big),
\label{eq:optimization}
\end{equation}
where $\mathrm{dist}\big( \cdot, \cdot \big)$ denotes the Euclidean distance between the eigenvalues of the two matrices, defined as
\begin{equation}
\mathrm{dist}\big( \boldsymbol{\Sigma}_h, \widehat{\boldsymbol{\Sigma}}_h \big) 
= \big\| \mathrm{Eig}(\boldsymbol{\Sigma}_h) - \mathrm{Eig}(\widehat{\boldsymbol{\Sigma}}_h) \big\|_2^2,
\label{eq:distance_metric}
\end{equation}
with $\mathrm{Eig} \left( \cdot \right )$ representing the vector of eigenvalues of the matrix. The overall algorithm that solves (\ref{eq:optimization}) is presented in \textbf{Algorithm 1}. The fundamental principle of \textbf{Algorithm 1} is to reformulate problem (\ref{eq:optimization}) as an eigenvalue allocation problem. Specifically, let $\Lambda_{2}=\{\lambda_{D+1},\ldots,\lambda_{N}\}$ denote the set of non-dominant eigenvalues (normalized to not exceed unity). These elements are processed sequentially: for the current eigenvalue $\lambda_{\mathrm{cur}}\in\Lambda_{2}$, we compute for each block $d\in\{1,\ldots,D\}$ the Euclidean distance metric $\mathrm{dist}_{d}$ between the provisional block–eigenvalue vector and the target correlation profile, and assign $\lambda_{\mathrm{cur}}$ to the block $d^{\star}=\arg\min_{d}\mathrm{dist}_{d}$. The procedure is repeated until all members of $\Lambda_{2}$ are placed, yielding a partition in which each block contains one dominant eigenvalue (from $\Lambda_{1}$) together with a set of subordinate eigenvalues (from $\Lambda_{2}$). The overall complexity of \textbf{Algorithm 1} is $\left(N-D\right) \times D$.

It is worth noting that two primary strategies exist for determining the correlation coefficient $\rho_d$, which can be described as follows:
\begin{itemize}
    \item \textbf{Constant Block-Correlation (CBC)}: In this method, a uniform correlation coefficient is applied across all blocks, while the block sizes $L_d$ are optimized to minimize the discrepancy between the dominant eigenvalues of $\boldsymbol{\Sigma}_h$ and $\widehat{\boldsymbol{\Sigma}}_h$, as formulated in (\ref{eq:optimization}). Specifically, $\rho_d$ is typically selected as a constant within the range $[0.95, 0.97]$ for all $d \in \left \{1,\ldots,D \right\}$. Although this approach provides a reasonable approximation for large $N$, it may lead to non-negligible approximation errors when $N$ is small, due to the sensitivity of the model to the choice of $\rho_d$.

    \item \textbf{Variable Block-Correlation (VBC)}: To overcome the limitations of the CBC model for systems with smaller $N$, the VBC model assigns a distinct correlation coefficient to each block. This enhanced flexibility improves the accuracy of the approximation. Moreover, closed-form expressions for this model have been derived and are given by \cite{ref17}
    \begin{equation}
\!\!\!\!\!\!\!\!\!\!\!\!\!{\rho}_{d}\!=\!\min \!\!\left(\!\max \!\left(\frac{\left(L_{d}-1\right) \lambda_{d}-\sum_{k=1}^{L_{d}-1} \lambda_{d, k}}{2\left(L_{d}-1\right)}, 0\right), 1\right).
\label{eq:VBC}
\end{equation}
\end{itemize}
In this paper, both correlation modeling strategies are incorporated into the analysis of the proposed FAS-RSMA framework, and their impact on OP and AC performance is thoroughly evaluated.
\vspace{-10pt}

\begin{algorithm}[t]
\caption{Block–Correlation Parameter Optimization for FAS–RSMA System}
\label{alg:VBC_opt}
\begin{algorithmic}[1]
\REQUIRE $D$, $\mathrm{Eig}(\boldsymbol{\Sigma}_h)=[\lambda_1,\ldots,\lambda_N]$
\STATE Split eigenvalues: $\Lambda_1=[\lambda_1,\ldots,\lambda_D]$, \;
       $\Lambda_2=[\lambda_{D+1},\ldots,\lambda_N]$
\STATE Initialize block eigenvalue sets $K_d \gets [\,]$ for $d=1,\ldots,D$
\FOR{$n=1$ \TO $N-D$}
  \STATE $\lambda_{\text{cur}} \gets \lambda_{D+n}$ and determine the $\rho_d^{\ast}$
  \FOR{$d=1$ \TO $D$}
    \STATE Compute potential error $\mathrm{dist}_d$ for assigning $\lambda_{\text{cur}}$ to block $d$
  \ENDFOR
  \STATE $d^{\ast} \gets \arg\min_{d}\, \mathrm{dist}_d$
  \STATE $K_{d^{\ast}} \gets [K_{d^{\ast}},\, \lambda_{\text{cur}}]$, $\rho_d\gets\rho_d^{\ast}$
\ENDFOR
\ENSURE Optimal parameters $\{L_d\}_{d=1}^{D}$ and $\{\rho_d\}_{d=1}^{D}$
\end{algorithmic}
\end{algorithm}

\subsection{Derivation of Outage Probability}
In this section, we present the derivation of a closed-form expression for the OP of the FAS-RSMA system. With the block-correlation framework, the original correlated channel matrix is transformed into $D$ independent blocks $\mathbf{A}_d$ defined in (\ref{eq:block_matrix_single}). For each block $d$, the $L_d$ spatially correlated channel coefficients can be represented using the decomposition of the constant correlation model, which is expressed as
\begin{equation}
\! \! \! \hat{h}_{i,d,u}\! \!  =\! \!  \sqrt{1\!\!  -\! \! \rho_d  } \left ( p_{1,i,d,u} \!\!  +\!\!   jp_{2,i,d,u}  \right ) \! \!  +\!\!  \sqrt{\! \rho_d} \left ( q_{1,d,u}\! \!   +\!\!   j q_{2,d,u}  \right ),
\label{eq:hatg}
\end{equation}
where $i \in \left \{ 1,\ldots, L_{d}\right \}$ and $\rho_d$ is determined by \textbf{Algorithm 1}. The random variables $p_{1,i,d,u}$, $p_{2,i,d,u}$, $q_{1,d,u}$, and $q_{2,d,u}$ are Gaussian variables with variance $\eta_0 /2$. The first complex term in (\ref{eq:hatg}) represents the independent channel component, while the second complex term accounts for the common component shared across all coefficients within block $d$. Consequently, the magnitude of $\hat{h}_{i,d,u}$ can be expressed as
\begin{equation}
\hat{g}_{i,d,u}=\sqrt{\sum_{t=1}^{2}\left(\sqrt{1-\rho_{d}} p_{t, i,d,u}+\sqrt{\rho_{d}} q_{t,d,u}\right)^{2}}.
\label{eq:gain}
\end{equation}
Leveraging the definition of the channel magnitude $\hat{g}_{i,d,u}$ in \eqref{eq:gain}, we state the following proposition.\\
\textbf{Proposition 3.} \textit{The joint conditional PDF of $\hat{\mathbf{g} }_{d,u}=\left \{ \hat{g}_{i,d, u} \mid i \in \left \{ 1, \ldots, L_{d} \right \}  \right \}$ for the $d$-th block is:
\begin{equation}
\begin{array}{l}
\! \! f_{\hat{\mathbf{g}}_{d,u} \mid \phi_{d}=\theta_{d}}\left(\mathbf{x}_{d,u}\right) \\
\! \! \! =\! \! \left(\! \frac{2}{\eta_{0}\left(1-\rho_{d}\right)}\! \right)^{\! \! L_{d}}\!  \prod_{i=1}^{L_{d}}\!  x_{i, d,u} e^{\! -\frac{x_{i, d,u}^{2}\! +\rho_{d} \theta_{d}^{2}}{\eta_{0}\left(1\! -\rho_{d}\right)}}\! \!  I_{0\! }\! \left(\! \frac{2 \sqrt{\rho_{d}} x_{i, d,u} \phi_{d}}{\eta_{0}\left(1-\rho_{d}\right)}\! \! \right)\! ,
\end{array}
\label{eq:p3}
\end{equation}
where $I_{0} \left(  \cdot \right )$ represents the zero-order modified Bessel function and $\mathbf{x}_{d,u}= \left[x_{1,d,u},\ldots, x_{L_d,d,u}  \right]$. The parameter $\phi_{d}$ is defined as
\begin{equation}
\phi_{d}=\sqrt{q_{1,d,u}^{2}+q_{2,d,u}^{2}}.
\label{eq:phi}
\end{equation}
}

\textit{Proof.} Starting from (\ref{eq:phi}), we condition on $\phi_{d}=\theta_d$, which implies that the random variable $\phi_d$ is fixed to a deterministic value. Under this condition, the random variable $\hat{g}_{i,d,u}$ can be interpreted as the magnitude of a complex Gaussian random variable with a deterministic mean shift $\mu_d=\sqrt{\rho_{d}} \theta_d$ introduced by the common component and variance $\sigma^2_d=\frac{\eta_{0} \left( 1-\rho_d \right)}{2}$. Consequently, $\hat{g}_{i,d,u}$ follows a Rician distribution, whose conditional PDF can be expressed as
\begin{equation}
\begin{array}{l}
f_{\hat{{g}}_{i,d,u} \mid \phi_{d} =\theta_{d}}\left({x}_{i,d,u}\right) \\
=\frac{2}{\eta_{0}\left(1-\rho_{d}\right)} x_{i, d,u} e^{-\frac{x_{i, d,u}^{2}+\rho_{d} \theta_{d}^{2}}{\eta_{0}\left(1-\rho_{d}\right)}} I_{0}\left(\frac{2 \sqrt{\rho_{d}} x_{i, d,u} \phi_{d}}{\eta_{0}\left(1-\rho_{d}\right)}\right).
\end{array}
\label{eq:rician}
\end{equation}
Since the correlation among channel coefficients within block $d$ is fully captured by the common term $q_{t,d,u}$, which is fixed once we condition on $\theta_d$, the channel coefficients within the block become conditionally independent. Therefore, the joint conditional PDF of the vector $\hat{\mathbf{g}}_{d,u}$ can be expressed as the product of the individual conditional PDFs:
\begin{equation}
f_{\hat{\mathbf{g}}_{d,u} \mid \phi_{d}=\theta_{d}}\left(\mathbf{x}_{d,u}\right) =\prod_{i=1}^{L_{d}}f_{\hat{{g}}_{i,d,u} \mid \phi_{d}=\theta_{d}}\left({x}_{i,d,u}\right).
\label{eq:joint}
\end{equation}
Finally, by substituting (\ref{eq:rician}) into (\ref{eq:joint}), we directly obtain the result in (\ref{eq:p3}), which proves the stated result.

Under circularly symmetric complex Gaussian fading—where the in-phase and quadrature components are i.i.d. $\mathcal{N}(0,\eta_0)$—the envelope $\phi_d$ follows a Rayleigh distribution. Its PDF is therefore given by
\begin{equation}
f_{\phi_{d}}\left( \theta_d \right) =\frac{2 \theta_d}{\eta_0}e^{-\frac{\theta^2_d}{\eta_0}}.
\label{eq:rayleigh}
\end{equation}
By utilizing \textbf{Proposition 3} in conjunction with (\ref{eq:rayleigh}), the unconditional PDF of $\hat{\mathbf{g}}_u=\left \{ \hat{\mathbf{g} }_{d,u} \mid d \in \left \{ 1, \ldots, D \right \}  \right \}$ can be expressed as
\begin{equation}
f_{\hat{\mathbf{g}}_{u}} \left ( \mathbf{x}_u  \right ) =\prod_{d=1}^{D} \int_{0}^{\infty } f_{\phi_{d}}\left( \theta_d \right)   f_{\hat{\mathbf{g}}_{d,u} \mid \phi_{d}=\theta_{d}}\left(\mathbf{x}_{d,u}\right) \mathrm{d}\theta_d,  
\label{eq:UcPDF}
\end{equation}
where $\mathbf{x}_u=\left \{ \mathbf{x}_{d,u} \mid d \in \left \{ 1, \ldots, D \right \}  \right \}$.

Based on \textbf{Proposition 1} and (\ref{eq:UcPDF}), we can now state the following theorem.\\
\textbf{Theorem 1:} \textit{The overall CDF of the $u$-th user's maximum channel amplitude is defined as
\begin{equation}
F_{\hat{{g}}_{u}}\!\! \left ( {x}_u  \right )\!\!=\!\!\prod_{d=1}^{D}\!\int_{0}^{\infty}\!\! \frac{2 \theta_d}{\eta_0}e^{-\frac{\theta^2_d}{\eta_0}} \!\left[ \!1\! -\! Q_{1} \!\!  \left( \!\frac{\theta_{d} \sqrt{\rho_{d}}}{\sigma_d}\!, \!\frac{x_{u}}{\sigma_d}\right )  \right]^{L_d} \!\!\!\!\mathrm{d}\theta_{d}.
\label{eq:cdf}
\end{equation}
The closed-form expression for the OP of the $u$-th user in the proposed FAS-RSMA system can be represented as
\begin{equation}
P_{\mathrm{out }}^{u} \! \!  \approx \! \! \prod_{d=1}^{D}\! \sum_{m=1}^{M} \omega_{m} \frac{2 \alpha_d}{\eta_{0}} e^{ \! -\frac{\alpha_d^{2}}{\eta_{0}}} \left[ \! 1 \!  -\!   Q_{1} \!\!  \left   ( \frac{\alpha_d\sqrt{\rho_{d}}}{\sigma_d}, \frac{\gamma_{\text {th }}^{u}}{\sigma_d}\right )  \right]^{L_d}\! \! ,
\label{eq:theorem1}
\end{equation}
where $\omega_{m}=H \pi \sqrt{1-t^{2}_{m}} /2M$ is the Gauss–Chebyshev weight factor with $H$ being the upper limit cutoff value \cite{lastone}, and $Q_{1}\left( \cdot,\cdot\ \!\right)$ denotes the first-order Marcum-$Q$ function \cite{lasttwo}. The parameters $\alpha_d$ and $t_m$ are defined as
\begin{subequations}\label{theorem1}
\begin{align}
 &\alpha_d=\frac{H}{2} \left(t_m+1 \right), \\
 &t_m=\cos \left (\frac{2m-1}{2M}  \pi \right ).
\end{align}
\end{subequations}
}

\textit{Proof:} See Appendix A for the proof.

The above theorem is derived under the assumption $\rho_d \in \left[0,1 \right)$. The boundary case $\rho_d = 1$ corresponds to \emph{perfect} intra-block correlation, where all ports within block $d$ are statistically identical and effectively collapse into a single antenna port. Hence, this regime is equivalent to the conventional traditional antenna system (TAS). Under $\rho_d = 1$, the PDF $\tilde{f}_{\hat{g}_{u}} \left( x_u \right)$ of channel gain $\hat{g}_{u}$ follows the Rayleigh distribution given by (\ref{eq:rayleigh}), and the corresponding OP for the block (equivalently, for TAS) is given by
\begin{align}
\tilde{P}_{out}^{u} 
=& \int^{\gamma_{\text {th }}^{u}}_0 \tilde{f}_{\hat{g}_{u}} \left( x_u \right) \mathrm{d}x_{u}\notag\\
=& \int^{\gamma_{\text {th }}^{u}}_0 \frac{2 x_{u}}{\eta_0}e^{-\frac{x^2_{u}}{\eta_0}}   \mathrm{d}x_u  \notag\\
=& 1-e^{ \! -\frac{(\gamma^{u}_{\text {th }})^{2}}{\eta_{0}}}.
\label{tas}
\end{align} 
\textbf{Remark 2:} \textit{The derived closed-form OP expression indicates a monotonic decrease in OP with both the SNR and the number of antenna ports $N$. This behavior is consistent with diversity principles: enlarging the port set enhances selection diversity through dynamic port reconfiguration, thereby improving reliability and yielding lower OP.}
\vspace{-10pt}

\subsection{Derivation of Achievable Capacity}
In this section, we present the derivation of a closed-form expression for the AC of the FAS-RSMA system. Based on \textbf{Theorem 1}, we can now derive the PDF of the $u$-th user's selected maximum channel amplitude as follows:
\begin{equation}
    \!\!f_{\hat{{g}}_{u}} \left ( x_u  \right )\!\! = \!\!\sum_{d=1}^{D}\!\! \left[ f^{*}_{\hat{{g}}_{d,u}} \!\!\left( x_u  \right)\!\!\!\!\! \prod_{j=1,j\ne d}^{D}\int_0^\infty \!\!  \!\!\hat{F}_{\hat{{g}}_{j,u}} \left ( {x}_u,\theta_j  \right ) \mathrm{d}\theta_j\right]\!,
\label{eq:pdfmax}
\end{equation}
where $\hat{F}_{\hat{{g}}_{j,u}} \left ( {x}_u,\theta_j  \right )$ is defined as
\begin{equation}
\hat{F}_{\hat{{g}}_{j,u}} \left ( {x}_u,\theta_j  \right )\!=\!\frac{2 \theta_j}{\eta_0}e^{-\frac{\theta^2_j}{\eta_0}} \!\left[ \!1\! -\! Q_{1} \!\!  \left( \!\frac{\theta_{j} \sqrt{\rho_{j}}}{\sigma_j}\!, \!\frac{x_{u}}{\sigma_j}\right )  \right]^{L_j}.    
\end{equation}
Meanwhile, the $f^{*}_{\hat{{g}}_{d,u}} \left( x_u  \right)=\frac{\mathrm{d}}{\mathrm{d}_{x_{u}}}F_{\hat{{g}}_{d,u}} \left ( {x}_u  \right )$, which can be expressed as
\begin{align}
f^{*}_{\hat{{g}}_{d,u}} \!\!\!\left( x_u  \right)\!\!
&=\! \!\!\int_0^\infty 
\!\!\frac{2 \theta_d}{\eta_0}e^{-\frac{\theta^2_d}{\eta_0}}\!\frac{\mathrm{d} }{\mathrm{d} x_u} 
\left\{\! \left[ 1\! - \!Q_1\!\!\left(\!\frac{\theta_d\sqrt{\rho_{d}}}{\sigma_d},\! \frac{x_u}{\sigma_d}\right) \right]^{L_d} \!\right\} 
\mathrm{d}\theta_d \notag \\
&=\!\!\! \int_0^\infty 
\!\!\frac{2 \theta_d}{\eta_0}e^{-\frac{\theta^2_d}{\eta_0}}L_d \left[ 1 \!- \!Q_1\left(\frac{\theta_d\sqrt{\rho_{d}}}{\sigma_d}, \frac{x_u}{\sigma_d}\right) \right]^{L_d-1}\notag \\ 
&\hspace{0.7cm} \times \frac{x_u}{\sigma^2_d}e^{-\frac{\left ( x^2_u+\rho_{d}\theta^2_d \right ) }{2 \sigma^2_d} }I_{0}\left ( \frac{\theta_d\sqrt{\rho_{d}} x_u}{\sigma^2_d}  \right ) \mathrm{d}\theta_d\notag \\ 
&=\!\!\! \int_0^\infty \hat{f}^{*}_{\hat{{g}}_{d,u}}\left(x_u,\theta_d \right)\mathrm{d}\theta_d.
\label{eq:pdff}
\end{align}
For the private message analysis, we have derived the PDF and CDF of the maximum channel amplitude for each user. However, in the case of the common message, the decoding performance is constrained by the user experiencing the weakest channel condition, i.e., the user with the smallest port amplitude. Therefore, it is essential to determine the statistical distribution of the minimum channel port amplitude among all the users, i.e., $f_{\hat{g}_{u^{*}}} \left(x^{*}_{u} \right)$. Based on this, we establish the following
proposition for the PDF of the minimum user channel gain $\hat{g}_{u^{*}}$.\\ \textbf{Proposition 4.} \textit{Assuming that all users share the same PDF $f_{\hat{{g}}_{u}}$ and CDF $F_{\hat{{g}}_{u}}$, and that the users are statistically independent, the PDF of $\hat{g}_{u^{*}}$ can be expressed as
\begin{equation}
f_{\hat{g}_{u^{*}}} \left( x_{u^{*}}  \right)=U\left [1- {F}_{\hat{g}_u}\left (x_{u^{*}}   \right )  \right ]^{U-1}{f}_{\hat{g}_u}\left (x_{u^{*}}   \right ).
\label{eq:commonpdf}
\end{equation}
}

\textit{Proof:} See Appendix B for the proof.

Building on \textbf{Proposition~2} and \textbf{Proposition~4}, substituting \eqref{eq:pdfmax} and \eqref{eq:commonpdf} into \eqref{eq:AC} yields \eqref{eq:long}, in which the AC of the common and private streams are expressed as double integrals with respect to $x_u$ and $\theta_j$. Exploiting the block–correlation structure and the resulting separability, we further obtain closed-form representations for both common and private streams, as stated in the following theorem.\\
\textbf{Theorem 2:} \textit{The closed-form expression for the AC of the common and private message can be represented as follows:
\begin{align}
{C}_c &\approx \sum_{l=1}^{M_l} \omega_{l} R_{c,u^{*}}  \left ( \alpha_d  \right )   \Big (1-\prod_{d=1}^{D} \sum_{s=1}^{M_s}\omega_{s} \hat{F}_{\hat{{g}}_{d,u}} \left (\alpha_d ,\beta _s  \right )     \Big )^{U-1}\notag \\
&\times \sum_{d=1}^{D}\Big[ \Big ( \sum_{s=1}^{M_s}\omega_{s} \hat{f}^{*}_{\hat{{g}}_{d,u}} \left (\alpha_d ,\beta _s  \right )  \Big ) \notag \\
&\times \prod_{j=1,j\ne d}^{D}\Big ( \sum_{s=1}^{M_s}\omega_{s} \hat{F}_{\hat{{g}}_{d,u}} \left (\alpha_d ,\beta _s  \right )  \Big ) \Big] .
\label{eq:theorem2c}
\end{align}
\begin{align}
{C}^{u}_p &\approx \sum_{l=1}^{M_l} \omega_{l} R_{p,u} \left ( \alpha_d  \right )\sum_{d=1}^{D}\Big[ \Big ( \sum_{s=1}^{M_s}\omega_{s} \hat{f}^{*}_{\hat{{g}}_{d,u}} \left (\alpha_d ,\beta _s  \right )  \Big ) \notag \\
&\times \prod_{j=1,j\ne d}^{D}\Big ( \sum_{s=1}^{M_s}\omega_{s} \hat{F}_{\hat{{g}}_{d,u}} \left (\alpha_d ,\beta _s  \right )  \Big ) \Big] .
\label{eq:theorem2p}
\end{align}
where $\omega_{l}=H_u \pi \sqrt{1-t^{2}_{l}} /2M$ and  $\omega_{s}=H_{\theta} \pi \sqrt{1-t^{2}_{s}} /2M$ are the Gauss–Chebyshev weight factors. The quadrature parameters are defined as follows:
\begin{equation}
 \begin{array}{ll}
t_{l}=\cos \left(\frac{(2 l-1) \pi}{2 M_{l}}\right), & \alpha_{d}=\frac{H_u}{2}\left(t_{l}+1\right), \\
t_{s}=\cos \left(\frac{(2 s-1) \pi}{2 M_{S}}\right), & \beta_{s}=\frac{H_{\theta}}{2}\left(t_{s}+1\right) .
\end{array}
\label{eq:quapara}
\end{equation}
}

\textit{Proof:} See Appendix C for the proof.
\begin{figure*}[t] 
\begin{align}
\bar{C}= &\int_{0}^{\infty }R_{c,u^{*}}U\left[1-F_{\hat{g}_u}\left(x_{u^{*}}\right)\right]^{U-1}\sum_{d=1}^{D}\left[\int_0^\infty \hat{f}^{*}_{\hat{{g}}_{d,u}}\left(x_{u^{*}},\theta_d \right)\mathrm{d}\theta_d\prod_{j=1,j\ne d}^{D}\int_{0}^{\infty}\!\!\!\hat{F}_{\hat{{g}}_{j,u}}\left(x_{u^{*}},\theta_j\right)\mathrm{d}\theta_j\right]\mathrm{d}x_{u^{*}}\notag \\
&+\sum^{u}_{u=1}\int_{0}^{\infty}R_{p,u}\sum_{d=1}^{D}\left[\int_0^\infty \hat{f}^{*}_{\hat{{g}}_{d,u}}\left(x_{u},\theta_d \right)\mathrm{d}\theta_d \prod_{j=1,j\ne d}^{D}\int_{0}^{\infty}\hat{F}_{\hat{{g}}_{j,u}}\left(x_u,\theta_j\right)\mathrm{d}\theta_j\right]\mathrm{d}x_u .
\label{eq:long}
\end{align}
\vspace{-1em}
\hrulefill
\end{figure*}

Similarly to \textbf{Theorem~1}, the above theorem is derived under the assumption $\rho_d \in \left[0,1 \right)$. For $\rho_d=1$, invoking \textbf{Proposition~4} and substituting \eqref{eq:rayleigh} together with \eqref{tas} into \eqref{eq:commonpdf}, the PDF of the worst-user gain $\hat{g}_{u^{*}}$ is obtained as
\begin{equation}
\tilde{f}_{\hat{g}_{u^{*}}} \left( x_{u^{*}}  \right)= U\times x_{u^{*}} e^{ \! -\frac{U \times(\gamma^{u}_{\text {th }})^{2}}{\eta_{0}}}.
\label{eq:commonpdf1}
\end{equation}
Following the same steps as in the proof of \textbf{Theorem~2}, we can have the following Corollary. \\
\textbf{Corollary 1:} \textit{The closed-form expressions for the AC of both the common and private messages for the case $\rho_d=1$ (equivalently, TAS) follow as
\begin{equation}
\tilde{C}_c \approx \sum_{l=1}^{M_l} \omega_{t} R_{c,u^{*}}  \left ( \alpha_t  \right )\tilde{f}_{\hat{g}_{u^{*}}} \left( \alpha_t   \right),
\label{eq:theorem2c1}
\end{equation}
\begin{equation}
\tilde{C}_p \approx \sum_{l=1}^{M_l} \omega_{t} R_{p,u} \left ( \alpha_t  \right )\tilde{f}_{\hat{g}_{u}} \left( \alpha_t   \right),
\label{eq:theorem2p1}
\end{equation}
where $\omega_{t}$ and $\alpha_t$ follow the same definition in (\ref{eq:quapara}). 
}\\
\textbf{Remark 3:} \textit{The closed-form AC expressions exhibit trends consistent with the OP analysis: the average capacity increases monotonically with SINR and with the number of FAS ports for both the common and private streams. These results highlight the practical benefit of integrating FAS with RSMA in system design.}

\section{Numerical Results and Analysis}
In this section, we present an extensive set of numerical results obtained under various simulation scenarios with different system configurations. The presented analysis encompasses both the verification of theoretical findings and the evaluation of the constant and variable block-correlation model performance, assessed across a range of system parameter settings.
\vspace{-15pt}

\subsection{System Parameters}
We normalize the average channel gain from the BS to all users as $\eta_0=1$, and the antenna array adopts a one-dimensional linear fluid antenna architecture with a maximum channel gain port selection scheme. The spatial correlation follows the Jakes model, as expressed in (\ref{Jakes}).

Furthermore, we focus on the performance evaluation of the FAS-RSMA system in a two-user scenario, noting that the extension to multiple users is straightforward. Unless stated otherwise, the power allocation factor for the common message is set to $t_c = 0.7$, while the power allocation coefficients for the private messages are given by $t_{p,1} = 0.6\left(1 - t_c \right)$ and $t_{p,2} = 0.4\left(1 - t_c \right)$, corresponding to user~1 and user~2, respectively. For the OP analysis, the parameters $\gamma^{u,c}_{\mathrm{th}}$ and $\gamma^{u,p}_{\mathrm{th}}$ are defined as $0~\text{dB}$ and $-6.5~\text{dB}$ for the common and private message thresholds, respectively. 

To ensure precise numerical evaluation of the derived analytical formulations, the Gauss-Chebyshev quadrature scheme settings are specified as follows: the integration limit is set to $H =H_u=H_{\theta} = 8\sqrt{\eta_0}$, and the number of quadrature nodes is $M =M_l=M_s = 30$. Monte Carlo simulations are performed with $N_{\mathrm{sim}}=1 \times 10^{6}$ realizations. 
\vspace{-10pt}

\subsection{Performance of Outage Probability}
In this section, the OP is employed as the performance metric to validate the accuracy of the derived analytical formulations and to facilitate comparison with various benchmark schemes for the FAS-RSMA system.

\begin{figure*}[!t]
\centering
\subfloat[$N=5, W=4$\label{fig:a}]{\includegraphics[width=0.38\linewidth]{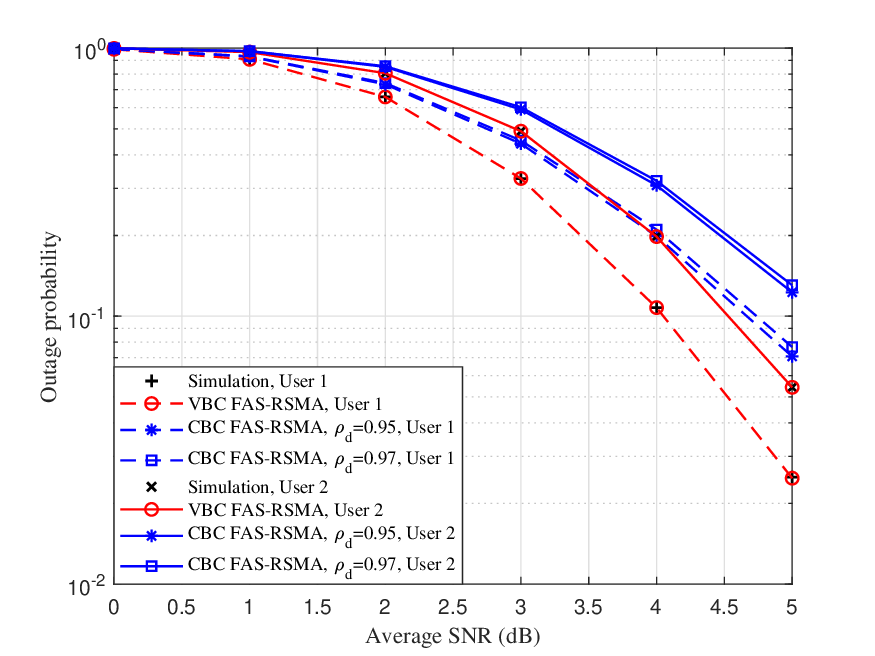}}
\hspace{0.02\linewidth}
\subfloat[$N=10, W=4$\label{fig:b}]{\includegraphics[width=0.38\linewidth]{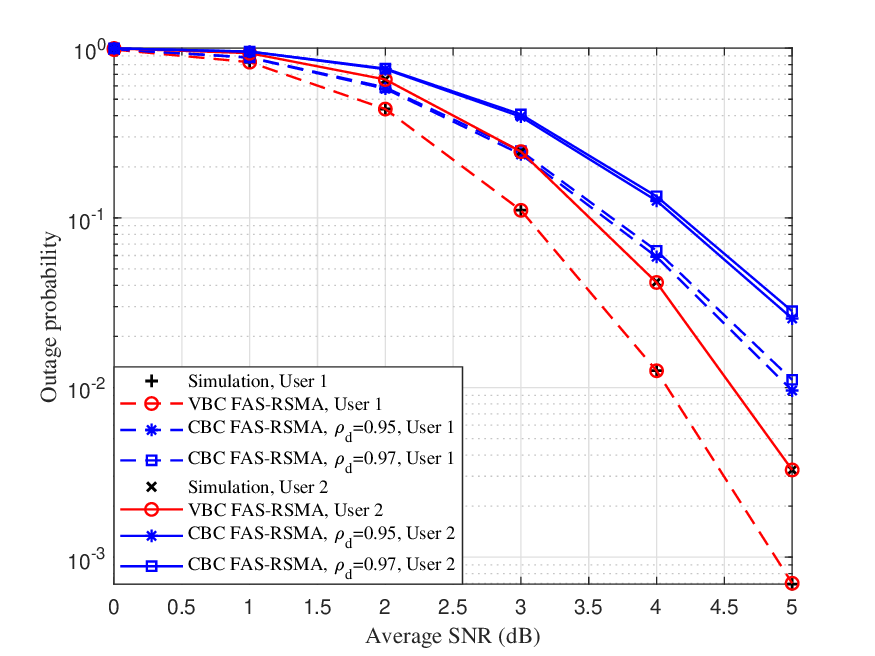}} \\[0.4em]
\subfloat[$N=5, W=8$\label{fig:c}]{\includegraphics[width=0.38\linewidth]{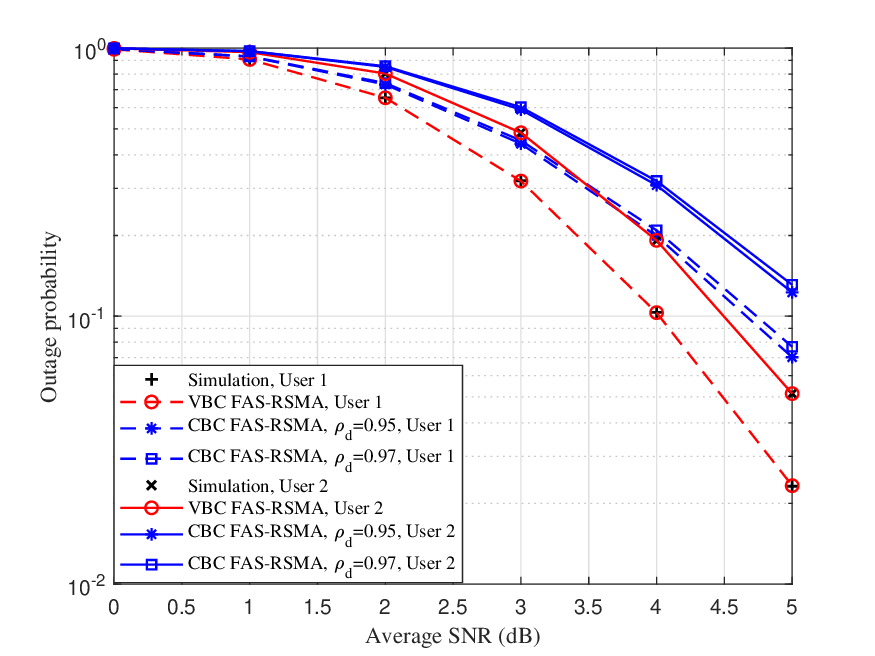}}
\hspace{0.02\linewidth}
\subfloat[$N=10, W=8$\label{fig:d}]{\includegraphics[width=0.38\linewidth]{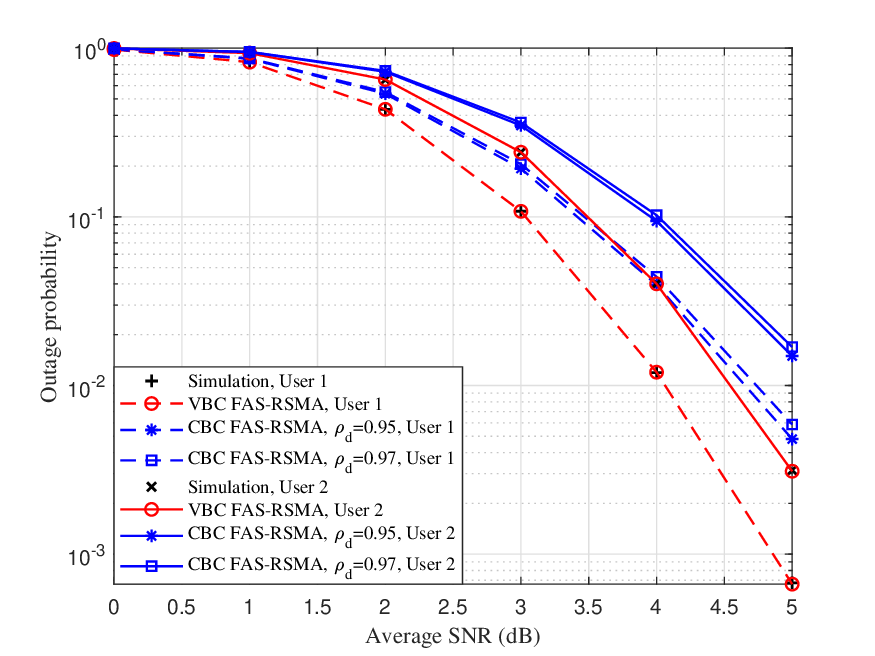}}
\caption{Outage probability comparison between different correlation models with different parameter configurations}
\vspace{-20pt}
\label{fig:OPVBC}
\end{figure*}
Fig.~\ref{fig:OPVBC} provides a detailed performance comparison between the VBC model and the CBC model for the FAS-RSMA system under different parameter settings, where the number of ports is varied as $N \in \left\{5, 10 \right\}$ and $W \in \left\{4, 8 \right\}$. Both analytical results (red lines for the VBC model and blue lines for the CBC model) and Monte Carlo simulation results (black markers) are provided. As illustrated in Fig.~\ref{fig:OPVBC}, it can be observed that the theoretical analysis of the VBC model exhibits excellent consistency with the Monte Carlo simulation compared to the CBC model for all system configurations. This close correspondence validates the closed-form OP expression derived in \textbf{Theorem 1}. Moreover, the discrepancy between the CBC model and the Monte Carlo simulation becomes increasingly evident at higher SNR values. The superior accuracy of the VBC model stems from its ability to represent the spatially nonuniform inter-port correlations intrinsic to FAS-based systems. By allowing blockwise variations in the correlation coefficients, VBC reproduces both dominant and subdominant eigenstructures of the channel covariance, thereby providing a closer approximation to the complex correlation patterns that arise in large antenna apertures and more accurately reflecting the selection-diversity gains of FAS. Even for compact configurations ($N = 5$), the VBC model achieves OP results that are in perfect agreement with the simulation outcomes, whereas the CBC model exhibits noticeable discrepancies. This consistency underscores the VBC model's robustness and confirms its applicability across a broad range of practical deployment scenarios. Additionally, a joint examination of the effects of $N$ and $W$ shows that increasing either parameter improves OP performance for both models and both users. A larger number of ports $N$ increases the port selection pool, thereby enhancing spatial diversity, while a larger normalization parameter $W$ expands the spatial sampling range, increasing the probability of encountering favorable channel realizations. Given the demonstrated superiority of the VBC model, all subsequent simulation analyses are conducted using the VBC model rather than the CBC counterpart.

\begin{figure}[!t]
\setlength{\belowcaptionskip}{-0.9cm} %
\vskip 0.1in
\begin{center}
\centerline{\includegraphics[width=0.43\textwidth]{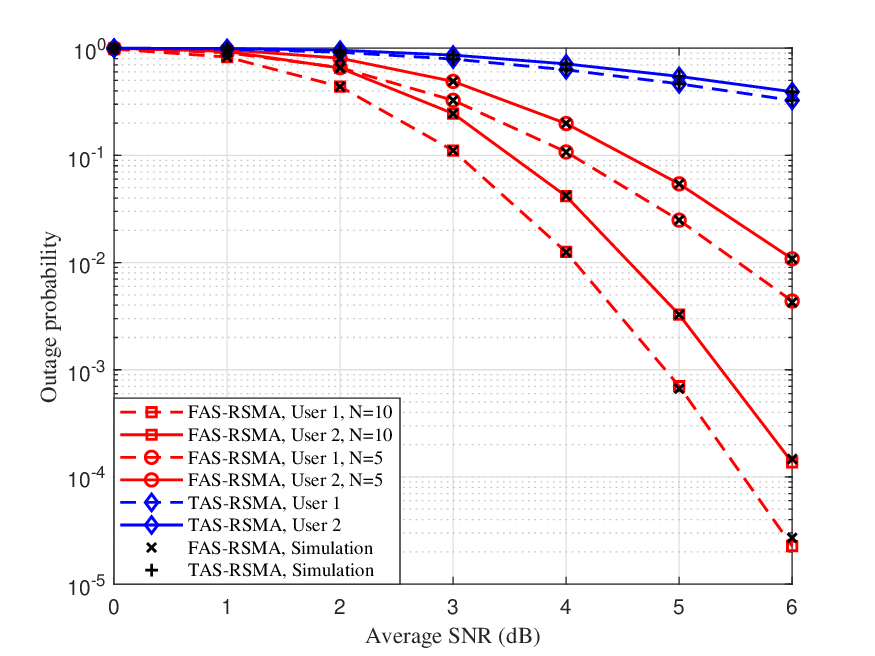}}
\caption{Performance of outage probability comparison between FAS and TAS of RSMA system.}
\vspace{-20pt}
\label{fig:OPTBC}
\end{center}
\vskip -0.1in
\end{figure}

Fig.~\ref{fig:OPTBC} depicts the OP performance of the proposed FAS-RSMA scheme in comparison with the conventional TAS-RSMA benchmark for different numbers of FAS ports. Both analytical results (solid and dashed lines) and Monte Carlo simulation results (markers) are provided for user~1 and user~2. First, it is evident that the simulation points align almost perfectly with the corresponding analytical curves for all configurations, thereby reaffirming the accuracy of the derived closed-form OP expression. Second, the proposed FAS-RSMA scheme consistently outperforms the conventional TAS-RSMA benchmark across the entire SNR range for both $N=5$ and $N=10$. The FAS-RSMA curves exhibit a noticeably steeper decay compared to TAS-RSMA, reflecting higher diversity gains. In particular, increasing the number of FAS ports from $N=5$ to $N=10$ yields a substantial reduction in OP for both users, especially in the high-SNR regime. This improvement arises from the enhanced spatial selectivity of FAS, which enables more efficient exploitation of spatial diversity and adaptive channel selection. Moreover, the performance gap between FAS-RSMA and TAS-RSMA widens as $N$ increases, indicating that the scalability of FAS can be effectively leveraged to further improve system robustness. Finally, another observation is that user~2 consistently experiences a slightly higher outage probability than user~1, which can be attributed to the asymmetric power allocation of the private messages for the two users in the RSMA framework. These results indicate that, relative to TAS-RSMA, integrating FAS into RSMA enhances spatial diversity, thereby yielding improved OP performance. To meet stringent reliability targets under practical SNR budgets, system deployments should prioritize FAS-RSMA with a sufficiently large switchable-port set, as this offers a scalable route to enhanced spatial diversity and reduced OP.

\begin{figure}[!t]
\setlength{\belowcaptionskip}{-0.9cm} %
\vskip 0.1in
\begin{center}
\centerline{\includegraphics[width=0.43\textwidth]{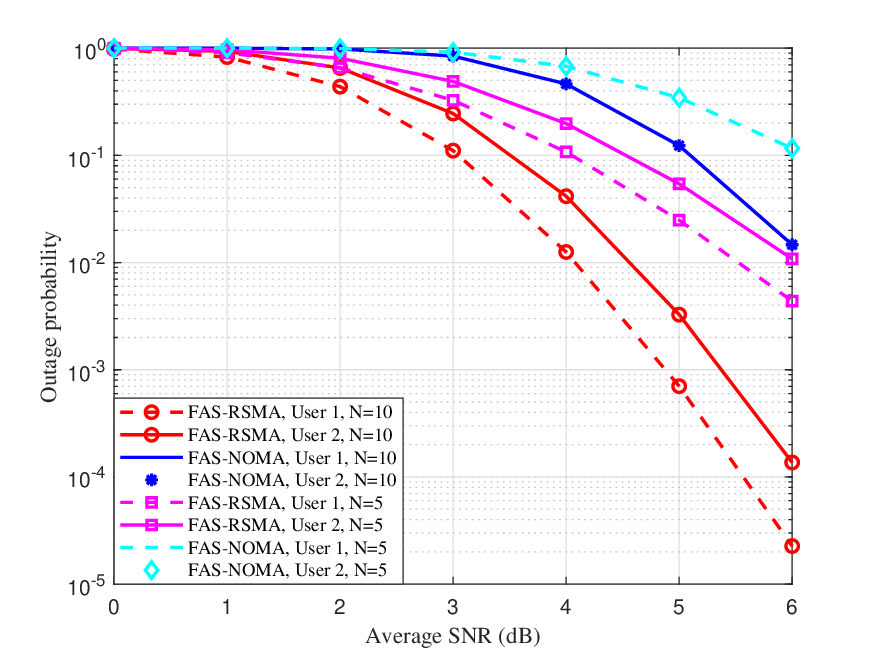}}
\caption{Performance of outage probability comparison between RSMA and NOMA.}
\vspace{-20pt}
\label{fig:OPNOMA}
\end{center}
\vskip -0.1in
\end{figure}

Fig.~\ref{fig:OPNOMA} compares the OP performance of FAS-RSMA and FAS-NOMA under different numbers of FAS ports. For each configuration, results are presented for both user~1 and user~2. For both $N$ values, the FAS-NOMA curves of user~1 and user~2 essentially overlap. This behavior stems from the current power allocation: the power split between the two NOMA users is not highly asymmetric, so although user~1 is assigned slightly more power, in the SIC order, both receivers first attempt to decode user~1's signal while still suffering a non-negligible interference term from user~2's signal. The resulting SINRs for that step are therefore similar at the two receivers, which yields nearly identical OPs. In contrast, FAS-RSMA assigns a sizable fraction of power to a common stream decoded by both users and then uses private streams for the residual information. Each receiver thus \emph{partially decodes} interference (via the common message) and \emph{partially treats} the remainder as noise with a flexible power split. This mechanism enhances the effective reliability for both users, yielding consistently lower OPs than FAS-NOMA across the entire SNR range. Meanwhile, increasing the number of ports from $N=5$ to $N=10$ improves both schemes due to stronger FAS port diversity; the improvement is more pronounced for FAS-RSMA, which exhibits a steeper high-SNR slope indicative of a higher diversity order. These findings suggest that under near-balanced user power allocation, NOMA's SIC provides little differentiation in robustness because both users face a similar interference-limited decoding condition in the first stage. RSMA overcomes this limitation by leveraging a power-boosted common stream and partial-interference decoding, thereby lowering both users' outage probabilities; larger $N$ further amplifies these gains via port diversity. Taken together, these findings point to a compelling design path that \emph{jointly} exploits FAS and RSMA. Under near-balanced user power budgets, FAS-RSMA should be favored over FAS-NOMA; allocate a nontrivial fraction of power to the common stream and provision a sufficiently large switchable-port set so that shared power and port diversity translate into systematically lower outage for both users.

\begin{figure*}[!t]
\centering
\subfloat[Common message power allocation factor $t_c$\label{fig:5a}]{\includegraphics[width=0.43\linewidth]{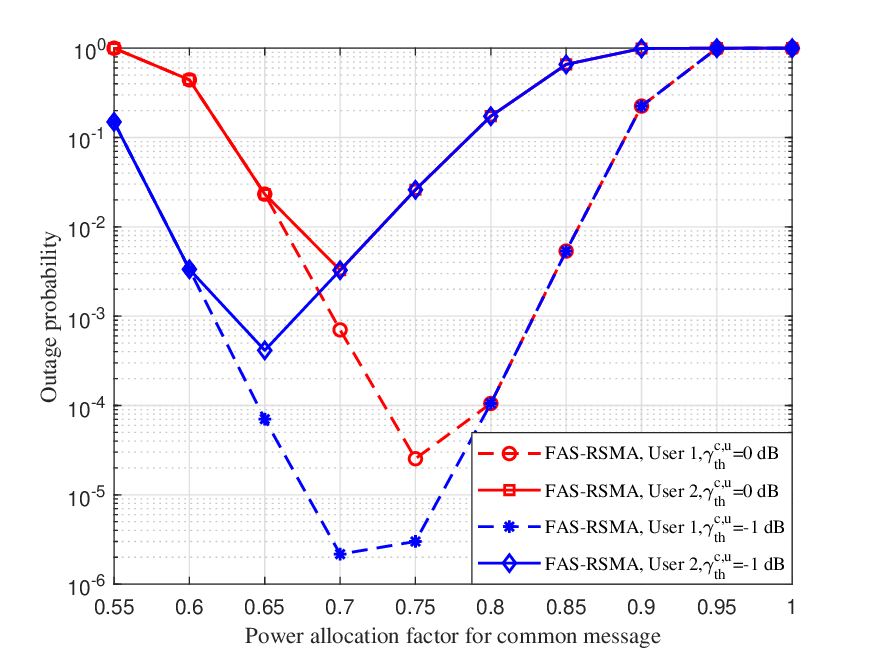}}
\hspace{0.02\linewidth}
\subfloat[Private message power allocation factor $t_{p,1}$\label{fig:5b}]{\includegraphics[width=0.43\linewidth]{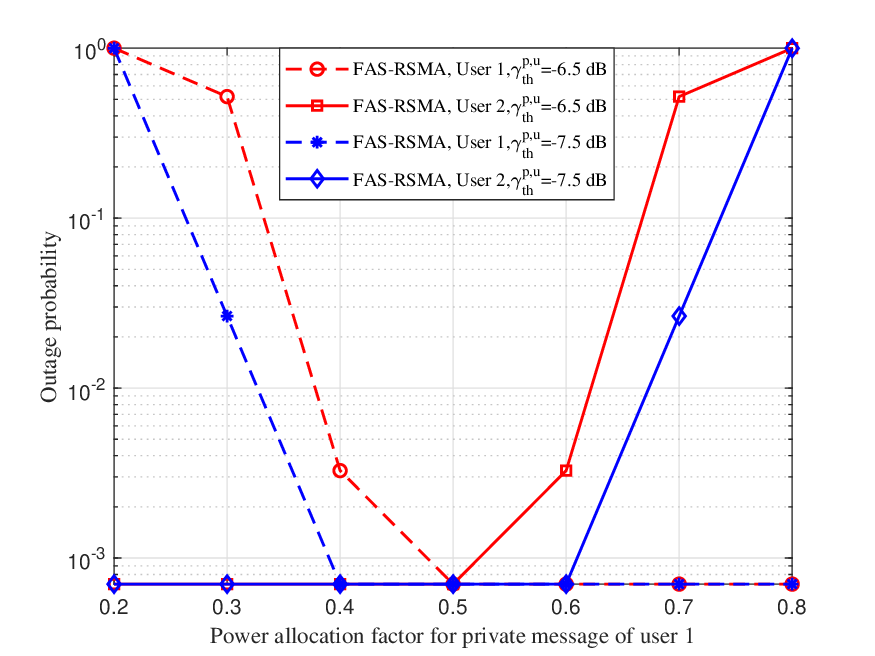}} 
\caption{Outage probability versus power allocation factors in FAS–RSMA. In (a), the private power fractions are tied to $t_{p,1}=0.6(1-t_c)$ and $t_{p,2}=0.4(1-t_c)$. In (b), the common power allocation is fixed to $t_c=0.7$ and the private power allocation varies.}
\vspace{-20pt}
\label{fig:powerallocation}
\end{figure*}

The impact of the message power allocation factor on OP performance is demonstrated in Fig.~5. Fig.~\ref{fig:powerallocation}(a) shows the OP as a function of the common power allocation factor $t_c$ under a fixed private power allocation factor $(t_{p,1},t_{p,2})=\big(0.6(1-t_c),\,0.4(1-t_c)\big)$. Two threshold values are considered: $\gamma^{c,u}_{\mathrm{th}}=-1~\mathrm{dB}$ (blue curves) and $\gamma^{c,u}_{\mathrm{th}}=0~\mathrm{dB}$ (red curves). For both threshold settings, the OP exhibits a convex-like dependence on the power allocation factor. When the allocation factor is small (below $0.6$), insufficient power is assigned to the common message, leading to high OP for both users. As the allocation factor increases, the OP decreases and reaches a minimum at an optimal point. Beyond this optimum, allocating more power to the common message degrades OP performance because the reduced power for the private messages increases the likelihood of private stream decoding failure. For $\gamma^{c,u}_{\mathrm{th}}=-1~\mathrm{dB}$, the optimal point $t^{\star}_c$ for OP performance is approximately $0.7$, where user~1 achieves OP between $10^{-5}$ and $10^{-6}$, and user~2 achieves OP down to $10^{-3}$. For $\gamma^{c,u}_{\mathrm{th}}=0~\mathrm{dB}$, the optimal allocation factor $t^{\star}_c$ shifts to around $0.75$, yielding OP as low as $10^{-4}$ and $10^{-2}$ for user~1 and user~2, respectively. This shift reflects the higher decoding requirement for the common message, which necessitates more power allocation to maintain reliability. From the results, we can observe that this convex-like trend arises from the inherent trade-off in distributing power between the common and private messages. The optimal value of the common message power allocation factor is related to the chosen outage threshold. Fig.~\ref{fig:powerallocation}(b) fixes $t_c=0.7$ and varies the private power allocation factor $t_{p,1}$. Two threshold values are considered: $\gamma^{p,u}_{\mathrm{th}}=-7.5~\mathrm{dB}$ (blue curves) and $\gamma^{p,u}_{\mathrm{th}}=-6.5~\mathrm{dB}$ (red curves). Similar to Fig.~\ref{fig:powerallocation}(a), the OP curves are convex-like: allocating too little power to one private stream (very small/large $t_{p,1}$) starves that user and raises the OP, whereas a balanced power split minimizes OP. The minimizing $t_{p,1}^\star$ depends on the private thresholds $\gamma^{p,u}_{\mathrm{th}}$: as $\gamma^{p,u}_{\mathrm{th}}$ increases (stricter decoding requirement), $t_{p,1}^\star$ moves toward a more even split (e.g., from the region $\left [0.4, 0.6 \right]$ at $\gamma^{p,u}_{\mathrm{th}}=-7.5$ dB toward $0.5$ at $\gamma^{p,u}_{\mathrm{th}}=-6.5$ dB), which balances both users' private SINR constraints under the fixed private budget $1-t_c=0.3$. These results provide a two-stage guideline for practical RSMA power design: first, select the common-stream fraction $t_c$ to satisfy the common decoding threshold with a safety margin, explicitly accounting for the tradeoff between common and private parts. Then, balance the private split near $t_{p,1}\!\approx\!t_{p,2}$ to jointly minimize both users' OP.

\vspace{-15pt}
\subsection{Performance of Average Capacity}
In this section, the AC is adopted as the performance metric to verify the accuracy of the derived analytical expressions and to enable a comparative evaluation for the FAS-RSMA system.

\begin{figure}[ht]
\setlength{\belowcaptionskip}{-0.9cm} %
\vskip 0.1in
\begin{center}
\centerline{\includegraphics[width=0.43\textwidth]{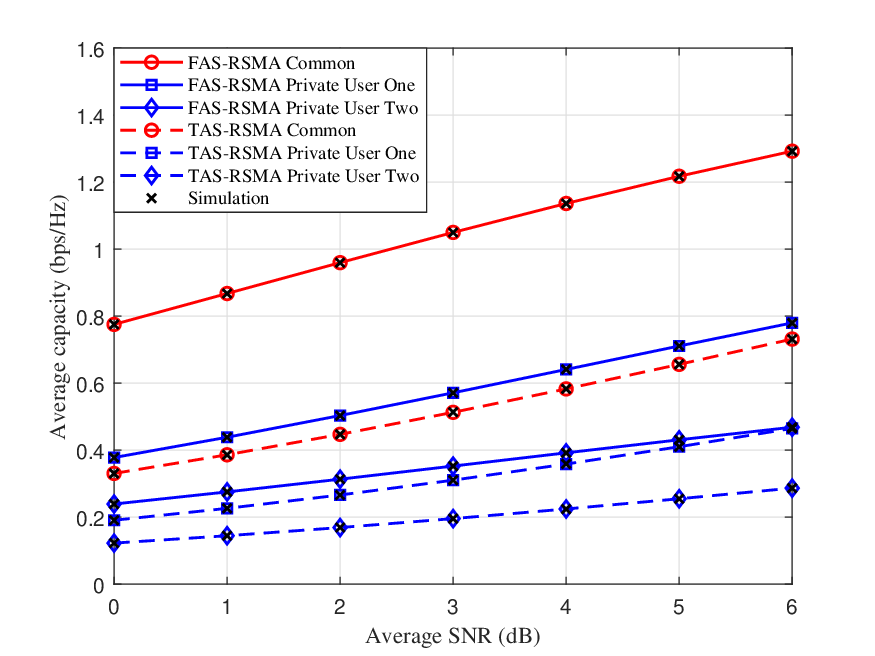}}
\caption{Performance of average capacity for common and private messages of FAS-RSMA and TAS-RSMA.}
\vspace{-15pt}
\label{fig:ACcommonpr}
\end{center}
\vskip -0.1in
\end{figure}

Fig.~\ref{fig:ACcommonpr} compares the average capacity performance of the common and private message parts for both TAS-RSMA and FAS-RSMA in the case of $N=5$ ports. The solid and dashed curves represent the analytical results, while the markers denote the results obtained from Monte Carlo simulations. The theoretical and simulation results exhibit an almost perfect match, thereby confirming the validity of the closed-form AC expressions for both the common and private messages derived in \textbf{Theorem~2} and \textbf{Corollary~1}. Furthermore, FAS-RSMA achieves superior performance over TAS-RSMA for both message types, owing to the dynamically reconfiguring ports of FAS, which enhance the effective channel quality. This widening gap is attributable to two factors: (i) RSMA allocates a nontrivial fraction of transmit power to the common stream, and (ii) the port-selection diversity inherent in FAS converts this additional common-stream power into a higher effective diversity gain. Consequently, the synergy between common-stream power boosting and FAS diversity amplifies the advantage of FAS over TAS at all SNR levels. These findings highlight that incorporating FAS into RSMA not only improves OP performance but also enhances AC for both the common and private message components. 

\begin{figure}[ht]
\setlength{\belowcaptionskip}{-0.9cm} %
\vskip 0.1in
\begin{center}
\centerline{\includegraphics[width=0.43\textwidth]{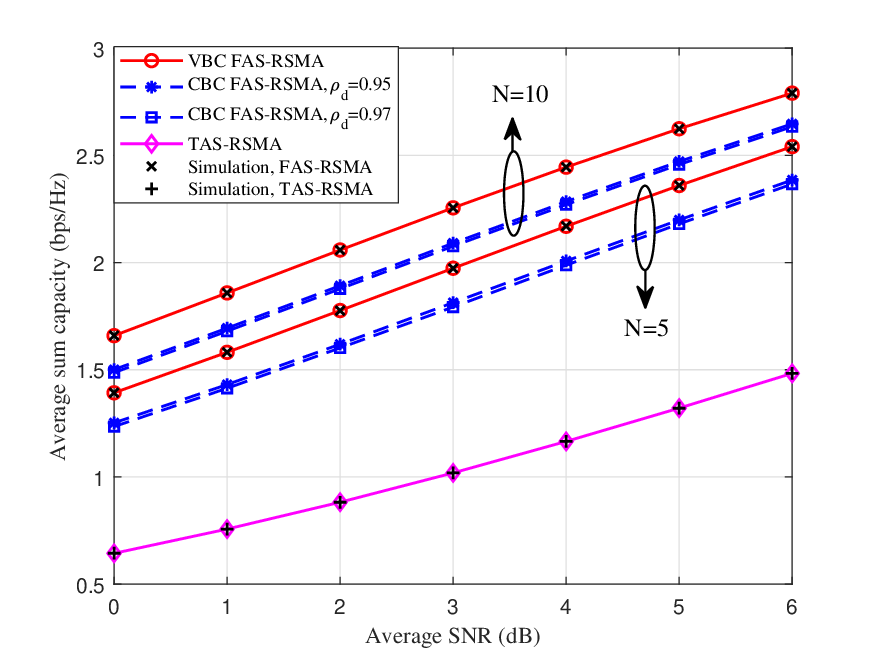}}
\caption{Performance of average sum capacity between different correlation models.}
\vspace{-20pt}
\label{fig:ACsum}
\end{center}
\vskip -0.1in
\end{figure}

Fig.~\ref{fig:ACsum} reports the average sum capacity of the proposed FAS-RSMA architecture under both the VBC and CBC models with $\rho_{d}\!\in\!\{0.95,0.97\}$ for two port counts $N\!\in\!\{5,10\}$; TAS-RSMA serves as a baseline. Across the entire SNR range and for both $N$ values, the simulation markers coincide almost perfectly with the VBC-based analytical curve. By contrast, the CBC-based curves display a systematic bias—most apparent at higher SNR and under stronger correlation—that tends to underestimate the achievable sum capacity. This discrepancy arises because VBC captures nonuniform inter-port correlation across the FAS aperture and thus represents selection diversity more accurately than the CBC model. Regardless of the correlation model, increasing the number of ports from $N=5$ to $N=10$ produces a near-uniform capacity uplift, confirming that reconfiguration diversity scales with $N$. Meanwhile, the TAS-RSMA baseline, which lacks port diversity, remains consistently lower AC compared with the FAS-RSMA curves, with its simulations matching the analysis closely. Collectively, these results indicate that reliable capacity prediction for FAS-RSMA necessitates correlation models that capture spatial heterogeneity (i.e., VBC), while enlarging the accessible port set monotonically enhances sum capacity; practically, this implies that correlation-aware (VBC-based) link budgeting and provisioning a larger switchable-port set should be prioritized in RSMA-enabled FAS designs.

\section{Conclusion}
This paper has developed a correlation-aware analytical framework for the FAS-RSMA system. FAS-driven dynamic port reconfiguration enables spatial selectivity for RSMA, strengthening the weakest user link and boosting SINR. Using block-correlation modeling with both CBC and VBC models, we derived closed-form expressions for OP and AC. Extensive simulations confirmed the analysis: VBC provided consistently tighter agreement with Monte Carlo simulation results than CBC, including regimes with limited port counts. Moreover, the synergy between RSMA's flexible interference management and FAS's dynamic port reconfiguration yielded clear improvements compared to traditional antenna systems and NOMA. The proposed framework offers practical guidance for correlation-aware transceiver design and constitutes a scalable basis for 6G terminals employing FAS-RSMA. Future work in FAS-RSMA will pursue several directions: (i) performance analysis for multi-antenna base stations with multiuser MIMO precoding; (ii) extensions to two-dimensional FAS apertures and mobility-induced correlation; (iii) robust transceiver design under imperfect CSI and hardware impairments; and (iv) joint optimization of port selection, power allocation, and rate-splitting parameters.

\vspace{-10pt}
\appendices
\section{Proof of Theorem 1}
Given \textbf{Proposition 1} and (\ref{eq:UcPDF}), the outage probability can be expressed as
\begin{align}
P_{\text{out}}^{d} 
&\overset{(a)}{=} \int_{0}^{\gamma^{u}_{\text{th}}} \cdots \int_{0}^{\gamma^{u}_{\text{th}}} 
f_{\hat{\mathbf{g}}_{u}} \left ( \mathbf{X}  \right )  \, \mathrm{d} x_{1,u} \cdots \mathrm{d}x_{N,u} \notag \\
&\overset{(b)}{=} \prod_{d=1}^{D} \int_{0}^{\infty} \frac{2 \theta_d}{\eta_0}e^{-\frac{\theta^2_d}{\eta_0}}\!\left[ 1 \!\!-\! Q_{1} \!\!  \left( \!\frac{\theta_{d}\! \sqrt{\rho_{d}}}{\sigma_d}, \frac{\gamma_{\text {th }}^{u}}{\sigma_d}\right )  \right]^{L_d} \!\!\!\mathrm{d}\theta_{d},
\label{eq:pout}
\end{align}
where step (a) follows directly from the definition of the CDF in (\ref{eq:P1}), step (b) exploits the fact that the channel variable $\hat{g}_{i,d,u}$ follows a Rician distribution under the fixed common component condition. Consequently, the conditional CDF of $\hat{g}_{i,d,u}$ can be expressed as
\begin{equation}
F_{\hat{{g}}_{i,d,u} \mid \phi_{d}=\theta_{d}} \left(x_{i,d,u} \right)\!\!=1 - Q_{1}  \left( \frac{\theta_{d} \sqrt{\rho_{d}}}{\sigma_d}, \frac{x_{i,d,u}}{\sigma_d}\right ),
\label{eq:cdfrician}
\end{equation}
where $Q_1(\cdot,\cdot)$ is the first-order Marcum $Q$-function. By substituting $x_{i,d,u}$ with $\gamma_{\text{th}}^u$, we obtain the CDF of the $u$-th user maximum channel port amplitude, as defined in \textbf{Theorem 1} (\ref{eq:cdf}).

To obtain a tractable closed-form approximation for the OP, the Gauss–Chebyshev quadrature is adopted for numerical evaluation. This method approximates an integral as a weighted summation of the function values evaluated at pre-defined nodes. The standard transformation for the Gauss–Chebyshev quadrature of the first kind is given by
\begin{equation}
\int_{-1}^{1} \frac{f(t)}{\sqrt{1-t^{2}}} d t \approx \frac{\pi}{N} \sum_{k=1}^{N_k} f\left(t_{k}\right),
\label{eq:gausschebyshev}
\end{equation}
where $t_k$ denotes the $k$-th Chebyshev node, which is defined as follows:
\begin{equation}
t_k=\cos \left( \frac{2k-1}{2N_k} \pi \right).
\label{eq:Chebyshevnode}
\end{equation}
By applying this quadrature to (\ref{eq:pout}), we perform the variable substitution $\theta_d=\frac{H}{2}(t_m+1)$, where $t_m\in[-1,1]$, to map the integration domain accordingly. The Jacobian of this transformation is given by $\frac{\mathrm{d}\theta_d}{\mathrm{d} t_m}=\frac{H}{2}$. This substitution transforms (\ref{eq:pout}) into (\ref{eq:theorem1}), which proves the stated result.
\vspace{-10pt}

\section{Proof of Proposition 4}
Since all the users are statistically independent, we can first find the CDF of the user with the smallest port amplitude $\hat{g}_{u^{*}}$. By the definition of choosing the minimum port amplitude among the users, we can write the following expression:
\begin{align}
    \Pr\{\hat{g}_{u^{*}} > x\} 
    &= \Pr\{\hat{g}_{1} > x,\, \hat{g}_{2} > x,\, \ldots,\, \hat{g}_{U} > x\} \notag \\
    &= \prod_{u=1}^U \Pr\{\hat{g}_{u} > x\} \notag \\
    &= \left[ 1 - F_{\hat{g}_u}(x) \right]^U .
    \label{eq:survival}
\end{align}
Consequently, the CDF of $\hat{g}_{u^{*}}$ can be expressed as
\begin{align}
    F_{\hat{g}_{u^{*}}}(x) 
    &= 1 - \Pr\{\hat{g}_{u^{*}} > x\} \notag \\
    &= 1 - \left[ 1 - F_{\hat{g}_u}(x) \right]^U .
    \label{eq:cdf_min}
\end{align}
To determine the PDF of $\hat{g}_{u^{*}}$, we need to differentiate $F_{\hat{g}_{u^{*}}}(x) $. Since $F_{\hat{g}_u}(x)$ is absolutely continuous, $F_{\hat{g}_{u^{*}}}(x)$ is differentiable almost everywhere. Differentiating \eqref{eq:cdf_min} with respect to $x$ yields
\begin{align}
    f_{\hat{g}_{u^{*}}}(x) 
    &= -\frac{d}{dx} \left[ 1 - F_{\hat{g}_u}(x) \right]^U \notag \\
    &= -U \left[ 1 - F_{\hat{g}_u}(x) \right]^{U-1} 
       \frac{d}{dx} \left[ 1 - F_{\hat{g}_u}(x) \right] \notag \\
    &= U \left[ 1 - F_{\hat{g}_u}(x) \right]^{U-1} f_{\hat{g}_u}(x),
\end{align}
which proves the stated result.
\vspace{-10pt}

\section{Proof of Theorem 2}
From the (\ref{eq:long}), it can be seen that it follows the double integrals structure for both common and private messages. Hence, we apply the Gauss-Chebyshev rule to approximate both inner and outer integrals.

Consider an integrable function $q$ on $[0,H]$. With the affine map
$z=\frac{H}{2}(t+1)$ sending $t\in[-1,1]$ to $z\in[0,H]$ and
$\mathrm{d}z=\frac{H}{2}\mathrm{d}t$, write
\begin{equation}
 \int_{0}^{H} q(z)\,\mathrm{d}z
= \int_{-1}^{1}\frac{\frac{H}{2}\sqrt{1-t^2}\,q\!\left(\frac{H}{2}(t+1)\right)}{\sqrt{1-t^2}}\,\mathrm{d}t.   
\end{equation}
Applying the Gauss-Chebyshev rule defined in (\ref{eq:gausschebyshev}) yields the mapped formula
\begin{equation}
\int_{0}^{H} q(z)\,\mathrm{d}z
\;\approx\; \sum_{k=1}^{N_k}\omega_k\,
q\!\Big(\tfrac{H}{2}(t_k+1)\Big),
\label{eq:GC-mapped}
\end{equation}
with Chebyshev nodes $t_k=\cos\!\big(\frac{(2k-1)\pi}{2N_k}\big)$.

\textbf{Inner $\theta$-integral. }
For each fixed $x$, apply \eqref{eq:GC-mapped} to every integral over
$\theta\in[0,H_\theta]$ with $M_s$ nodes, then we have
\begin{equation}
\omega_s=\tfrac{H_\theta\pi}{2M_s}\sqrt{1-t_s^2},\quad
t_s=\cos\!\Big(\tfrac{(2s-1)\pi}{2M_s}\Big),    
\end{equation}
which gives
\begin{align}
\int_{0}^{H_\theta}\!\hat F_{\hat g_{j,u}}(x_{u^{*}},\theta_j)\,\mathrm{d}\theta_j
&\approx \sum_{s=1}^{M_s}\omega_s\,\hat F_{\hat g_{j,u}}(x_{u^{*}},\beta_s),\\
\int_{0}^{H_\theta}\!\hat f^{*}_{\hat g_{d,u}}(x_u,\theta_j)\,\mathrm{d}\theta_j
&\approx \sum_{s=1}^{M_s}\omega_s\,\hat f^{*}_{\hat g_{d,u}}(x_u,\beta_s),
\end{align}
where $\beta_s=\tfrac{H_\theta}{2}(t_s+1)$ denotes the $s$-th Gauss-Chebyshev node for the inner integral. Hence, every product of inner integrals factorizes into products of Gauss-Chebyshev sums, which is shown in (\ref{eq:long2}).

\begin{figure*}[t] 
\begin{align}
\bar{C}\approx &\int_{0}^{\infty }R_{c,u^{*}}U\left[1-\prod_{d=1}^{D} \sum_{s=1}^{M_s}\omega_{s} \hat{F}_{\hat{{g}}_{d,u}} \left (x_u^{*},\beta_s  \right )  \right]^{U-1}\sum_{d=1}^{D}\left[\sum_{s=1}^{M_s}\omega_s\,\hat f^{*}_{\hat g_{d,u}}(x_u^{*},\beta_s)\prod_{j=1,j\ne d}^{D}\sum_{s=1}^{M_s}\omega_s\,\hat F_{\hat g_{j,u}}(x_{u^{*}},\beta_s)\!\right]\!\mathrm{d}x_{u^{*}}\notag \\
&+\sum^{u}_{u=1}\int_{0}^{\infty}R_{p,u}\sum_{d=1}^{D}\left[\sum_{s=1}^{M_s}\omega_s\,\hat f^{*}_{\hat g_{d,u}}(x_u,\beta_s) \prod_{j=1,j\ne d}^{D}\sum_{s=1}^{M_s}\omega_s\,\hat F_{\hat g_{j,u}}(x_{u^{*}},\beta_s)\right]\mathrm{d}x_u .
\label{eq:long2}
\end{align}
\vspace{-1em}
\hrulefill
\end{figure*}

\textbf{Outer $x$-integral.}  
After replacing all inner integrals by the Gauss-Chebyshev sums above, shown in (\ref{eq:long2}), apply
\eqref{eq:GC-mapped} to the remaining outer integral over $x_u^{*}, x_u \in[0,H_x]$ with
$M_l$ nodes, then we have
\begin{equation}
\omega_l=\tfrac{H_u\pi}{2M_l}\sqrt{1-t_l^2},\quad
t_l=\cos\!\Big(\tfrac{(2l-1)\pi}{2M_l}\Big).    
\end{equation}
which gives
\begin{align}
\int_{0}^{H_u} R_{c,u^{*}}\Psi(x_{u^{*}},\beta_s)\,\mathrm{d}x_{u^{*}}
&\approx \sum_{l=1}^{M_l}\omega_l\,R_{c,u^{*}} (\alpha_d)\,\Psi(\alpha_d,\beta_s),\\
\int_{0}^{H_u} R_{p,u} \Phi(x_u,\beta_s)\,\mathrm{d}x_u
&\approx \sum_{l=1}^{M_l}\omega_l\,R_{p,u}(\alpha_d)\,\Phi(\alpha_d,\beta_s),
\label{eq:outer}
\end{align}
where $\alpha_d=\tfrac{H_u}{2}(t_l+1)$ denotes the $l$-th Gauss-Chebyshev node for the outer integral, and the functions $\Psi(x_{u^{*}},\beta_s)$ and $\Phi(x_u,\beta_s)$ collect the remaining integral terms in \eqref{eq:long2}. Substituting \eqref{eq:outer} into \eqref{eq:long2} then yields the quadrature-based discrete representations in \eqref{eq:theorem2c} and \eqref{eq:theorem2p}, thereby establishing the stated result.


 




\vfill


\begin{thebibliography}{1}
\bibliographystyle{IEEEtran}

\bibitem{ref1}
W. K. New \emph{et al.}, ``A Tutorial on Fluid Antenna System for 6G Networks: Encompassing Communication Theory, Optimization Methods and Hardware Designs," {\it{IEEE Commun. Surv. Tutor.}}, early access, doi:10.1109/COMST.2024.3498855, 2024.

\bibitem{ref2}
T. Wu, \emph{et al.}. ``Fluid antenna systems enabling 6G: Principles, applications, and research directions." {\it{arXiv preprint}}, arXiv:2412.03839, 2024.


\bibitem{ref20}
T. Han and K. Kobayashi, ``A new achievable rate region for the interference channel," {\it{IEEE Trans. Inf. Theory}}, vol. 27, no. 1, pp. 49–60, 1981.

\bibitem{ref21}
B. Clerckx, H. Joudeh, C. Hao, M. Dai, and B. Rassouli, ``Rate splitting for MIMO wireless networks: a promising PHY-layer strategy for LTE evolution,"  {\it{IEEE Commun. Mag.}}, vol. 54, no. 5, pp. 98–105, 2016.

\bibitem{ref210}
H. Joudeh and B. Clerckx, “Sum-Rate Maximization for Linearly Precoded Downlink Multiuser MISO Systems with Partial CSIT: A Rate-Splitting Approach,” {\it{IEEE Trans. on Comm.}}, vol. 64, no. 11, pp. 4847-4861, Nov 2016.

\bibitem{ref2105}
Y. Mao, B. Clerckx, and V. O. K. Li, “Rate-splitting multiple access for downlink communication systems: bridging, generalizing, and outperforming SDMA and NOMA,” {\it{EURASIP J. Wireless Commun. Netw.}}, vol. 2018, no. 1, p. 133, May 2018.

\bibitem{ref211}
B. Clerckx, Y. Mao, R. Schober, and H. V. Poor, “Rate-splitting unifying SDMA, OMA, NOMA, and multicasting in MISO broadcast channel: A simple two-user rate analysis,” {\it{IEEE Wireless Commun. Lett.}}, vol. 9, no. 3, pp. 349–353, Nov. 2019.

\bibitem{ref212}
A. Bansal, K. Singh, B. Clerckx, C.-P. Li, and M.-S. Alouini, ``Rate-splitting multiple access for intelligent reflecting surface aided multi-user communications," {\it{IEEE Trans. Veh. Technol.}}, vol. 70, no. 9, pp. 9217–9229, Sep. 2021.

\bibitem{ref22}
Y. Mao, O. Dizdar, B. Clerckx, R. Schober, P. Popovski, and H. V. Poor, ``Rate-Splitting Multiple Access: Fundamentals, Survey, and Future Research Trends," {\it{IEEE Commun. Surveys Tuts.}}, vol. 24, no. 4, pp. 2073–2126, 2022.

\bibitem{ref221}
A. Mishra, Y. Mao, O. Dizdar, and B. Clerckx, “Rate-splitting multiple access for downlink multiuser MIMO: Precoder optimization and PHY layer design,” {\it{IEEE Trans. Commun.}}, vol. 70, no. 2, pp. 874–890, Feb. 2022.

\bibitem{ref23}
B. Clerckx \emph{et al.}, ``A Primer on Rate-Splitting Multiple Access: Tutorial, Myths, and Frequently Asked Questions," {\it{IEEE J. Sel. Areas Commun.}}, vol. 41, no. 5, pp. 1265–1308, 2023.

\bibitem{ref232}
J. Liu \emph{et al.}, ``Energy efficiency of rate-splitting multiple access for multibeam satellite communications," in {\it{Proc. IEEE 97th Veh. Technol. Conf. (VTC)}}, 2023, pp. 1–5.

\bibitem{ref233}
X. Lyu, S. Aditya, J. Kim, and B. Clerckx, “Rate-splitting multiple access: The first prototype and experimental validation of its superiority over SDMA and NOMA,” {\it{IEEE Trans. Wireless Commun.}}, vol. 23, no. 8, pp. 9986-10000, Aug. 2024

\bibitem{ref234}
S. Zhang, B. Clerckx, D. Vargas, O. Haffenden, and A. Murphy, “Rate-splitting multiple access: Finite constellations, receiver design, and SIC-free implementation,” {\it{IEEE Trans. Commun.}}, vol. 72, no. 9, pp. 5319-5333, Sept. 2024

\bibitem{ref24}
J. Liu \emph{et al.}, ``Power allocation for high mobility OTFS-RSMA system with path-selective one-tap receiver," {\it{IEEE Wireless Commun. Lett.}}, vol. 14, no. 3, pp. 661-665, March 2025.

\bibitem{ref241}
S. K. Singh, K. Agrawal, K. Singh, and C.-P. Li, ``Outage probability and throughput analysis of UAV-assisted rate-splitting multiple access," {\it{IEEE Wireless Commun. Lett.}}, vol. 10, no. 11, pp. 2528–2532, Nov. 2021.


\bibitem{ref243}
Y. Tong \emph{et al.}, ``Outage analysis of rate splitting networks with an untrusted user," {\it{IEEE Trans. Veh. Technol.}}, vol. 72, no. 2, pp. 2626–2631, Feb. 2023.

\bibitem{ref244}
F. Xiao, X. Li \emph{et al.}, ``Outage performance analysis of RSMA-aided semi-grant-free transmission systems," {\it{IEEE Open J. Commun. Soc.}}, vol. 4, pp. 253–268, 2023.


\bibitem{ref245}
M. Dai, B. Clerckx, D. Gesbert, and G. Caire, “A rate splitting strategy for massive MIMO with imperfect CSIT,” {\it{IEEE Trans. Wireless Commun.}}, vol. 15, no. 7, pp. 4611–4624, Jul. 2016.

\bibitem{ref246}
Y. Mao, B. Clerckx, J. Zhang, V. O. Li, and M. A. Arafah, “Max-min
fairness of K-user cooperative rate-splitting in MISO broadcast channel
with user relaying,” {\it{IEEE Trans. Wireless Commun.}}, vol. 19, no. 10,
pp. 6362–6376, Oct. 2020.



\bibitem{ref4}
K. K. Wong, A. Shojaeifard, K.-F. Tong, and Y. Zhang, ``Performance limits of fluid antenna systems," {\it{IEEE Commun. Lett.}}, vol. 24, no. 11, pp. 2469–2472, Nov. 2020.

\bibitem{ref5}
K.-K. Wong \emph{et al.}, ``Fluid antenna system," {\it{IEEE Trans. Wireless Commun.}}, vol. 20, no. 3, pp. 1950-1962, Mar 2021.

\bibitem{ref18}
P. Ram´ırez-Espinosa, D. Morales-Jimenez, and K. K. Wong, ``A new spatial block-correlation model for fluid antenna systems," {\it{IEEE Trans. Wireless Commun.}}, vol. 23, no. 11, pp. 15829–15843, Nov. 2024.

\bibitem{ref26}
Y. Shen \emph{et al.}, ``Design and implementation of mmWave surface wave enabled fluid antennas and experimental results for fluid antenna multiple access," {\it{arXiv preprint}}, arXiv:2405.09663, May 2024.

\bibitem{ref27}
B. Liu, K.-F. Tong, K. K. Wong, C.-B. Chae, and H. Wong, ``Programmable meta-fluid antenna for spatial multiplexing in fast fluctuating radio channels," {\it{Opt. Express}}, vol. 33, no. 13, pp. 28898–28915, 2025.

\bibitem{ref28}
J. Zhang \emph{et al.}, ``A novel pixel-based reconfigurable antenna applied in fluid antenna systems with high switching speed," {\it{IEEE Open J. Antennas Propag.}}, vol. 6, no. 1, pp. 212–228, Feb. 2025.

\bibitem{ref6}
K.-K. Wong and K.-F. Tong, ``Fluid antenna multiple access," {\it{IEEE Trans. Wireless Commun.}}, vol. 21, no. 7, pp. 4801-4815, Jul. 2022.

\bibitem{ref7}
S. Li, \emph{et al.},``Outage analysis of NOMA-enabled backscatter communications with intelligent reflecting surfaces," {\it{IEEE Internet Things J.}}, vol. 9, no. 16, pp. 15390-15400, Aug. 2022

\bibitem{ref8}
F. R. Ghadi \emph{et al.}, ``Copula-based performance analysis for fluid antenna systems under arbitrary fading channels," {\it{IEEE Commun. Lett.}}, vol. 27, no. 11, pp. 3068–3072, Nov. 2023.

\bibitem{ref84}
Q. Zhang \emph{et al.}, “An efficient sum-rate maximization algorithm for fluid antenna-assisted ISAC system,” {\it{IEEE Commun. Lett.}}, vol. 29, no. 1, pp. 200–204, Jan. 2025.

\bibitem{ref85}
H. Niu \emph{et al.}, ``A Survey on Artificial Noise for Physical Layer Security: Opportunities, Technologies, Guidelines, Advances, and Trends, '' \emph{ IEEE Commun. Surv. Tutor. }, 2025, early access, doi: 10.1109/COMST.2025.3610758.

\bibitem{ref9}
W. K. New, K. K. Wong, H. Xu, K. F. Tong and C.-B. Chae, ``Fluid antenna system: New insights on outage probability and diversity gain," {\it{IEEE Trans. Wireless Commun.}}, vol. 23, no. 1, pp. 128–140, Jan. 2024.

\bibitem{ref10}
X. Lai \emph{et al.}, ``On performance of fluid antenna system using maximum ratio combining," {\it{IEEE Commun. Lett.}}, vol. 28, no. 2, pp. 402–406, Feb. 2024.

\bibitem{ref11}
H. Xu \emph{et al.}, ``Channel estimation for FAS-assisted multiuser mmWave systems," {\it{IEEE Commun. Lett.}}, vol. 28, no. 3, pp. 632–636, Mar. 2024

\bibitem{ref110}
W. Kiat New \emph{et al.}, "Channel Estimation and Reconstruction in Fluid Antenna System: Oversampling is Essential," {\it{IEEE Trans. Wireless Commun.}}, vol. 24, no. 1, pp. 309-322, Jan. 2025.

\bibitem{ref12}
W. K. New \emph{et al.}, ``An information-theoretic characterization of MIMO-FAS: Optimization, diversity-multiplexing tradeoff and q-outage capacity," {\it{IEEE Trans. Wireless Commun.}}, vol. 23, no. 6, pp. 5541–5556, Jun. 2024.

\bibitem{ref13}
J. Yao \emph{et al.}, ``Exploring Fairness for FAS-assisted Communication Systems: from NOMA to OMA," {\it{IEEE Trans. Wireless Commun.}}, vol. 24, no. 4, pp. 3433-3449, April 2025.

\bibitem{ref14}
C. Wang \emph{et al.}, ``Fluid antenna system liberating multiuser MIMO for ISAC via deep reinforcement learning," {\it{IEEE Trans. Wireless Commun.}}, vol. 23, no. 9, pp. 10879-10894, Sep. 2024.

\bibitem{ref15}
L. Zhou \emph{et al.}, ``Fluid antenna-assisted ISAC systems," {\it{IEEE Wireless Commun. Lett.}}, vol. 13, no. 12, pp. 3533– 3537, Dec. 2024

\bibitem{ref16}
J. Yao \emph{et al.}, ``FAS-RIS communication: Model, analysis, and optimization," {\it{IEEE Trans. Veh. Technol.}}, vol. 74, no. 6, pp. 9938-9943, June 2025.


\bibitem{ref17}
X. Lai \emph{et al.}, ``Revisiting spatial block-correlation model for fluid antenna systems: from constant to variable correlations," {\it{arXiv preprint}}, 2024.

\bibitem{ref19}
X. Lai \emph{et al.}, ``FAS-RIS: A block-correlation model analysis," {\it{IEEE Trans. Veh. Technol.}}, vol. 74, no. 2, pp. 3412-3417, Feb. 2025.


\bibitem{ref250}
Zhang, Cixiao \emph{et al.} ``Fluid Antenna-Aided Robust Secure Transmission for RSMA-ISAC Systems," {\it{arXiv preprint}}, 2025.

\bibitem{ref25}
F. R. Ghadi\emph{et al.} ``Fluid antenna-aided rate-splitting multiple access," {\it{IEEE Trans. Veh. Technol.}}, 2025, early access, doi: 10.1109/TVT.2025.3599709.

\bibitem{ref29}
F. R. Ghadi \emph{et al.} ``Phase-mismatched STAR-RIS with FAS-assisted RSMA Users," {\it{arXiv preprint}}, 2025.

\bibitem{ref30}
F. R. Ghadi \emph{et al.} ``UAV-Relay Assisted RSMA Fluid Antenna System: Outage Probability Analysis," {\it{arXiv preprint}}, 2025.

\bibitem{lastone}
E. Suli and D. F. Mayers, {\it{An Introduction to Numerical Analysis.}} Cambridge, U.K.: Cambridge Univ. Press, 2003

\bibitem{lasttwo}
I. S. Gradshteyn and I. M. Ryzhik, {\it{Table of Integrals, Series, and Products}}, 7th ed. San Diego, CA: Academic, 2007.
\end{thebibliography}
\end{document}